\documentclass[twocolumn,secnumarabic,amssymb, nobibnotes, aps, prc, superscriptaddress,nobalancelastpage,nofootinbib]{revtex4-1}
\usepackage{bm} 
\usepackage{amsmath} \usepackage{braket} \usepackage{epsfig}
\usepackage{tensor}
\usepackage[version=4]{mhchem}
\usepackage{titlesec}
\usepackage{ifthen}
\usepackage{sidecap}
\usepackage{listings}
\usepackage[para,online,flushleft]{threeparttablex}
\usepackage{mathtools}
\usepackage{siunitx}
\usepackage[normalem]{ulem}
\usepackage{natbib}
\usepackage{multirow}
\usepackage{enumitem}
\usepackage{subfig}
\usepackage{graphicx}
\usepackage{booktabs}
\usepackage{physics}

\usepackage{xr-hyper}
\usepackage{hyperref}
\usepackage[all]{hypcap}
\usepackage[dvipsnames]{xcolor}
\hypersetup{breaklinks=true,colorlinks=true,linkcolor=blue,citecolor=blue,filecolor=magenta,urlcolor=cyan}

\def\ref@jnl#1{{\jnl@style#1}}

\def\aj{\ref@jnl{AJ}}                   
\def\actaa{\ref@jnl{Acta Astron.}}      
\def\araa{\ref@jnl{ARA\&A}}             
\def\apj{\ref@jnl{ApJ}}                 
\def\apjl{\ref@jnl{ApJ}}                
\def\apjs{\ref@jnl{ApJS}}               
\def\ao{\ref@jnl{Appl.~Opt.}}           
\def\apss{\ref@jnl{Ap\&SS}}             
\def\aap{\ref@jnl{A\&A}}                
\def\aapr{\ref@jnl{A\&A~Rev.}}          
\def\aaps{\ref@jnl{A\&AS}}              
\def\azh{\ref@jnl{AZh}}                 
\def\baas{\ref@jnl{BAAS}}               
\def\bac{\ref@jnl{Bull. astr. Inst. Czechosl.}}
\def\caa{\ref@jnl{Chinese Astron. Astrophys.}}
\def\cjaa{\ref@jnl{Chinese J. Astron. Astrophys.}}
\def\icarus{\ref@jnl{Icarus}}           
\def\jcap{\ref@jnl{J. Cosmology Astropart. Phys.}}
\def\jrasc{\ref@jnl{JRASC}}             
\def\memras{\ref@jnl{MmRAS}}            
\def\mnras{\ref@jnl{MNRAS}}             
\def\na{\ref@jnl{New A}}                
\def\nar{\ref@jnl{New A Rev.}}          
\def\pra{\ref@jnl{Phys.~Rev.~A}}        
\def\prb{\ref@jnl{Phys.~Rev.~B}}        
\def\prc{\ref@jnl{Phys.~Rev.~C}}        
\def\prd{\ref@jnl{Phys.~Rev.~D}}        
\def\pre{\ref@jnl{Phys.~Rev.~E}}        
\def\prl{\ref@jnl{Phys.~Rev.~Lett.}}    
\def\pasa{\ref@jnl{PASA}}               
\def\pasp{\ref@jnl{PASP}}               
\def\pasj{\ref@jnl{PASJ}}               
\def\rmxaa{\ref@jnl{Rev. Mexicana Astron. Astrofis.}}%
\def\qjras{\ref@jnl{QJRAS}}             
\def\skytel{\ref@jnl{S\&T}}             
\def\solphys{\ref@jnl{Sol.~Phys.}}      
\def\sovast{\ref@jnl{Soviet~Ast.}}      
\def\ssr{\ref@jnl{Space~Sci.~Rev.}}     
\def\zap{\ref@jnl{ZAp}}                 
\def\nat{\ref@jnl{Nature}}              
\def\iaucirc{\ref@jnl{IAU~Circ.}}       
\def\aplett{\ref@jnl{Astrophys.~Lett.}} 
\def\apspr{\ref@jnl{Astrophys.~Space~Phys.~Res.}}
\def\bain{\ref@jnl{Bull.~Astron.~Inst.~Netherlands}}
\def\fcp{\ref@jnl{Fund.~Cosmic~Phys.}}  
\def\gca{\ref@jnl{Geochim.~Cosmochim.~Acta}}   
\def\grl{\ref@jnl{Geophys.~Res.~Lett.}} 
\def\jcp{\ref@jnl{J.~Chem.~Phys.}}      
\def\jgr{\ref@jnl{J.~Geophys.~Res.}}    
\def\jqsrt{\ref@jnl{J.~Quant.~Spec.~Radiat.~Transf.}}
\def\memsai{\ref@jnl{Mem.~Soc.~Astron.~Italiana}}
\def\nphysa{\ref@jnl{Nucl.~Phys.~A}}   
\def\physrep{\ref@jnl{Phys.~Rep.}}   
\def\physscr{\ref@jnl{Phys.~Scr}}   
\def\planss{\ref@jnl{Planet.~Space~Sci.}}   
\def\procspie{\ref@jnl{Proc.~SPIE}}   

\newcommand{\nmp}{nuclear matter parameters }

\newcommand{\nsat}{n_{sat}}
\newcommand{\esat}{E_{sat}}
\newcommand{\ksat}{K_{sat}}
\newcommand{\qsat}{Q_{sat}}
\newcommand{\zsat}{Z_{sat}}
\newcommand{\lsym}{L_{sym}}
\newcommand{\esym}{E_{sym}}
\newcommand{\ksym}{K_{sym}}
\newcommand{\qsym}{Q_{sym}}
\newcommand{\zsym}{Z_{sym}}
\newcommand{\ncc}{n_{cc}}
\newcommand{\lk}{\mathcal{L}}

\begin{document}

\title{Properties of the neutron star crust informed by nuclear structure data}

\author{Pietro Klausner}
\email{pietro.klausner@unimi.it}
\affiliation{Universit\'e de Caen Normandie, CNRS/in2p3, LPC Caen (UMR6534), 14050 Caen, France}
\affiliation{Dipartimento di Fisica ``Aldo Pontremoli'', Universit\`a degli Studi di Milano, 20133 Milano, Italy}
\affiliation{INFN, Sezione di Milano, 20133 Milano, Italy}

\author{Marco Antonelli}
\email{antonelli@lpccaen.in2p3.fr}
\affiliation{Universit\'e de Caen Normandie, CNRS/in2p3, LPC Caen (UMR6534), 14050 Caen, France}

\author{Francesca Gulminelli}
\email{gulminelli@lpccaen.in2p3.fr}
\affiliation{Universit\'e de Caen Normandie, CNRS/in2p3, LPC Caen (UMR6534), 14050 Caen, France}

\begin{abstract}
We perform a Bayesian analysis of the neutron star (NS) equation of state (EoS) based on a wide set of Skyrme functionals, derived from previous nuclear physics inferences. The novelty of this approach lies in starting from the full multidimensional posterior distribution of nuclear matter parameters, consistent with a comprehensive set of static and dynamic nuclear structure observables. We construct unified EoSs for $npe\mu$ matter, where the inner crust of the NS is treated using an extended Thomas-Fermi method, providing for the first time a fully consistent Bayesian treatment of the correlation of bulk with surface as well as with spin-orbit and effective mass parameters. We then employ a standard Bayesian framework to identify those EoSs that satisfy astrophysical constraints from NS mass measurements, the tidal deformability from GW170817, and NICER mass-radius observations. We also examine NS observables, such as the crustal moment of inertia, which is crucial in understanding pulsar glitches. Compared to previous works, we observe an increase in both the NS surface thickness and the crustal moment of inertia.
\end{abstract}

\maketitle

\section{Introduction}
\label{sec:intro}

The equation of state (EoS) of neutron stars (NSs) remains a major uncertainty in nuclear astrophysics, as a wide range of baryonic densities must be covered, which cannot be accessed by a single theory or experimental approach.  
While the high-density behavior of dense matter is mostly probed by astrophysical observations, nuclear physics experiments provide robust constraints at sub-saturation densities. 
A key challenge is to ensure that nuclear uncertainties are properly accounted for when extrapolating empirical information on finite nuclei to deduce the bulk behavior of matter.

In this work, we take a Bayesian approach to the NS EoS, leveraging a wide set of Skyrme functionals derived from a previous inference using nuclear structure data~\cite{Klausner2025}.  
Unlike traditional studies that impose nuclear matter parameter priors in an ad hoc manner~\cite{Tsang2024}, our approach starts from the full posterior distribution of Skyrme functionals that are consistent with a large set of static and dynamical nuclear experimental observables.  
These include masses and charge radii, spin-orbit splittings, the excitation energies of the isoscalar giant monopole and quadrupole resonances, the energy-weighted sum rule of the isovector giant dipole, as well as the electric dipole polarizabilities $\alpha_D$ of $^{208}$Pb and $^{48}$Ca~\cite{Tamii2011, Birkhan2017}, and the parity-violating asymmetries $A_{PV}$ measured in the PREX-II and CREX experiments~\cite{PREX-2,CREx},see~\cite{Klausner2025} for details.  
These latter observables are sensitive to the neutron-skin thickness of nuclei, which is deeply linked to the symmetry energy and the EoS of pure neutron matter~\cite{Reinhard2022}.  
Moreover, it has been argued that both $\alpha_D$ and $A_{PV}$ measurements entail relevant consequences for NSs~\cite{Koliogiannis2025, Reed2021, Fattoyev2018}.  
Deriving the prior from an experiment-informed posterior enables an exact treatment of correlations between nuclear matter parameters~\cite{XRM2015}, while preserving the information content of the nuclear observables and providing a more physically grounded prior for the~NS~EoS.

From these Skyrme functionals, we construct unified EoSs for $npe\mu$ matter, where the inner crust is treated using an extended Thomas-Fermi method that makes full use of the nuclear-physics-informed finite-size terms of the energy functional.  
The extension to densities above saturation is achieved using the meta-modeling technique, while maintaining consistency with the underlying functional form.  
We then perform a Bayesian inference to determine which of these unified EoSs satisfy astrophysical constraints, including NS mass measurements, tidal deformability constraints from GW170817, and NICER mass-radius observations.  
Additionally, we explore key NS observables, such as the crustal moment of inertia, which plays a crucial role in pulsar glitch models~\cite{amp_review_2023}.

The paper is structured as follows. In Section~\ref{sec:meta}, we briefly present the meta-modeling technique and explain how we compute the NS EoS from it. 
Section~\ref{sec:bayes} is dedicated to describing our Bayesian setup and how we map the Skyrme posterior into a MM prior. 
Section~\ref{sec:crust} presents the numerical details of the computation of the inner crust. We then report our main results in Section~\ref{sec:res}, and conclude in Section~\ref{sec:concl} with a summary and some perspectives.

\section{Stellar EoS from the  nuclear Skyrme-Meta-Model}
\label{sec:meta}

The metamodel (MM) introduced in~\citet{metamodel2018} provides a relatively simple analytic representation of the nuclear EoS at zero temperature and for a given proton fraction, under the core assumption that matter is purely nucleonic.
For astrophysical applications, it is complemented with electrons and muons in weak equilibrium, forming cold-catalyzed $npe\mu$ matter. 
In this study, we enhance the bulk functional given by the MM with the finite-size terms from the Skyrme interaction. 
This enables us to use the full multidimensional posterior from the comprehensive Skyrme inference of~\citet{Klausner2025} as a nuclear-informed prior in our analysis.

The free parameters of the analytic representation are the so-called nuclear matter parameters (NMP), i.e., the density derivatives at saturation of the energy per particle of symmetric nuclear matter and of the symmetry energy. 
These parameters are traditionally grouped into \emph{isoscalar} ($\nsat$, $\esat$, $\ksat$) and \emph{isovector}~($\esym$, $\lsym$, $\ksym$~\cite{Chen2009}.  

In addition, two parameters are introduced to impose the values of the effective masses of neutrons and protons in pure neutron matter and symmetric matter at saturation ($m^*_{IS}$ and $m^*_{IV}$), as well as a phenomenological constant $b = 10\ln(2)$, which ensures the correct behavior of the MM at very low densities~\cite{metamodel2018}.
These bulk parameters are complemented by the standard finite-size terms from the Skyrme interaction~\cite{Chabanat1997}, namely, one spin-orbit parameter and two surface parameters associated with the isoscalar and isovector gradient terms. 
The choice of this parameter set enables a one-to-one mapping between the MM and the Skyrme functional.

To complete the MM representation of the nuclear functional for homogeneous matter, four additional parameters are introduced: $\qsat$, $\zsat$, $\qsym$, and $\zsym$, which correspond to higher orders of the usual Taylor expansion of the energy per nucleon around saturation~\cite{metamodel2018}. 

Starting from a set of bulk and surface parameters, we compute the $\beta$-equilibrated EoS and the star composition as a function of the baryon densities $n_B$, with a unified treatment of the core and the crust. 
The inner crust EoS and composition are treated within an approximated extended Thomas-Fermi (ETF) technique ~\cite{Pearson2018, Shchechilin2024}, detailed in Section~\ref{sec:crust}.
We start from $n_B=2.66\times10^{-4}$ fm$^{-3}$, the neutron drip density for the BSk24 functional~\cite{BSk24}, as we do not compute the composition nor the EoS of the outer crust, but rather use those of the BSk24 energy density functional 
\footnote{The outer crust data was taken from the CompOSE online repository~\cite{ compOSE}.}. 
The outer crust is composed of neutron-rich nuclei, and one could, in principle, leverage the information coming from our previous nuclear inferences.
However, the shell effects, which are predominant there due to the absence of the free neutron gas, cannot be accounted for by the ETF formalism, and one should use full quantum approaches like HFB instead.
Although these methods are more precise, they come at a heavier computational cost, which is not compatible with a Bayesian approach like ours.
The extended Skyrme BSk24 was fit to a comprehensive data set spanning the whole nuclear chart, and, therefore, we expect it to be the best likelihood model of a hypothetical Bayesian analysis constrained by that data set.
Furthermore, the uncertainty in the EOS of the outer crust due to the uncertainty of the nuclear model is too small to have a noticeable effect on the global star observables we will compute \cite{Pearson2018}.
That is why we feel confident that using BSk24 outer crust EOS will not appreciably affect our results.

\section{Bayesian Inference}
\label{sec:bayes}

In this work, we update the previous results of \citet{Klausner2025} through a new Bayesian inference, with neutron star observations as constraints. 
This is done swiftly in a Bayesian framework by using the posterior distribution obtained in \cite{Klausner2025} as this work's prior distribution, thus naturally incorporating the information on nuclear experiments.

Still, employing a standard Skyrme interaction as the base model for the nuclear EoS would be unwise, as its lack of flexibility at super-saturation densities would hinder our ability to describe neutron star properties.
As far as homogeneous EoS properties are concerned, the Skyrme parameter set allows to independently vary only five \nmp ($n_{sat}$, $\esat$, $\ksat$, $\esym$, and $\lsym$); see~\cite{Chen2009}.
The higher order derivatives $Q_{sat,sym}$, $Z_{sat,sym}$, and $\ksym$, crucial for describing nuclear matter at high-density, are therefore not free parameters but rather a function of those five.

Passing to the MM representation solves this problem, since we can decouple the high-order parameters from the lower-order ones and remove the limitations imposed by the Skyrme functional form.
The five free \nmp ($n_{sat}$, $\esat$, $\ksat$, $\esym$ and $\lsym$), as well as the two effective masses $m^*_{IS}$ and $m^*_{IV}$ (isoscalar and isovector), are mapped exactly in their MM counterparts; $\ksym$, pivotal for the sound speed, will be computed following the Skyrme formulae \cite{Dutra2012}.
While this does not decouple $\ksym$ from the other parameters, it ensures the continuity of the sound speed in the star medium.
As for $Q_{sat,sym}$ and $Z_{sat,sym}$, we use the Skyrme ones if the baryon density is below saturation; otherwise, we employ randomly picked values. 
This way, we keep the behavior of the Skyrme equations of state at low densities, where they were fitted, while allowing more freedom at high densities.

\begin{figure}
    \includegraphics[width=1\linewidth]{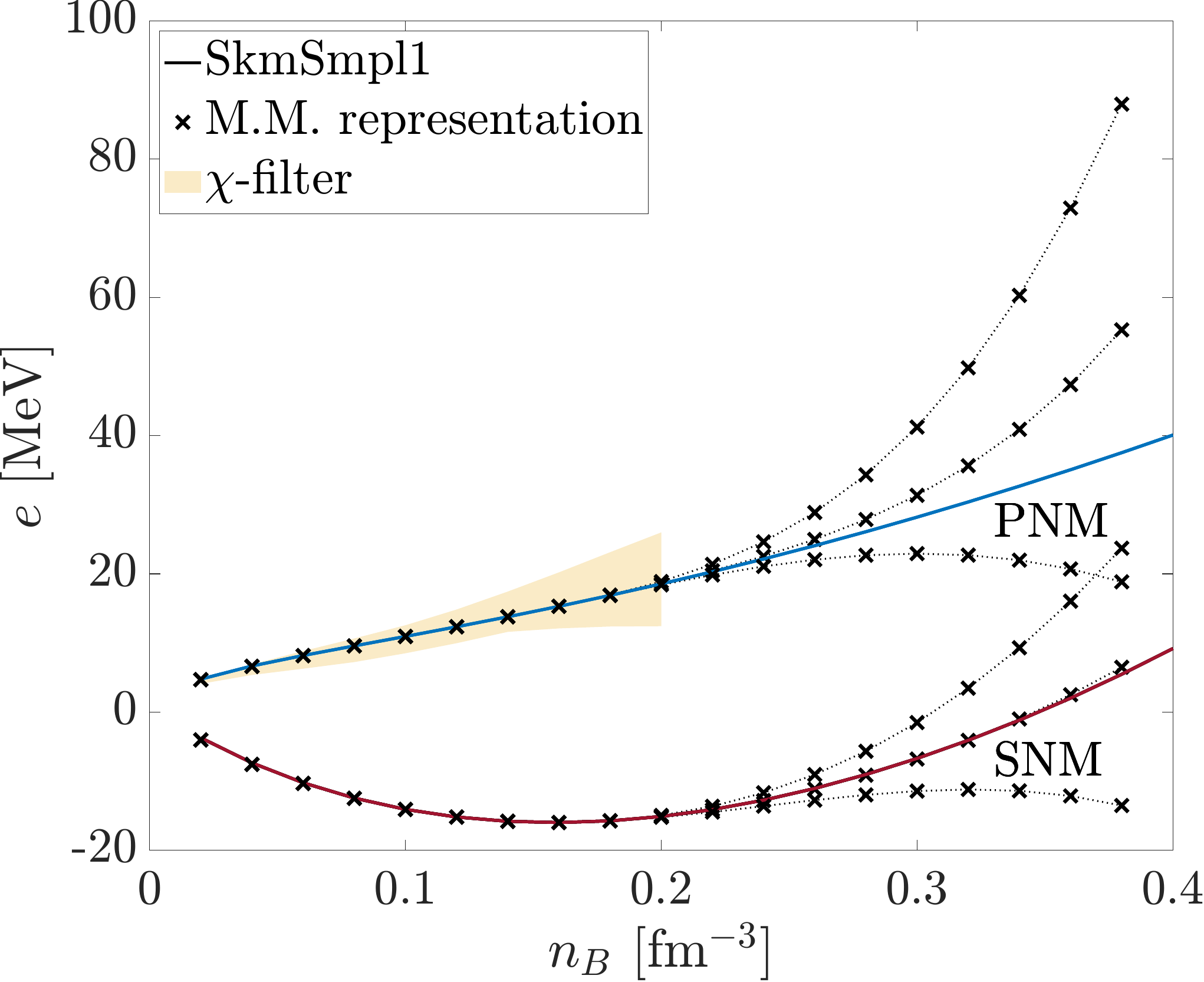}
    \caption{Mapping of SkmSmpl1 EoS (line) into three different possible realizations of the MM (black crosses) for symmetric nuclear matter (SNM, in red) and pure neutron matter (PNM, in blue); see text for details.
    The yellow region delimits the 90\% interval of the region predicted by ab-initio calculations employing chiral interactions, taken from~\cite{Huth2021}.}
    \label{fig:mapping}
\end{figure}

To illustrate the procedure, Figure~\ref{fig:mapping} shows three possible MM representations for a randomly selected Skyrme interaction from \cite{Klausner2025}, which we call ``SkmSmpl1''. 
Its parameters are given in appendix~\ref{app:sampled_skyrme}. 
The solid line represents the Skyrme nuclear EoS, in red for symmetric nuclear matter (SNM) and blue for pure neutron matter (PNM). 
The black crosses are the corresponding quantities computed with the three MM representations of SkmSmpl1, as prescribed above. 
Up until 0.2 fm$^{-3}$ they are almost identical (indeed, they are the same below $\nsat$ as we employ SkmSmpl1 $Q_{sat,sym}$ and $Z_{sat,sym}$) and faithfully reproduce the Skyrme EoS; at higher densities, they diverge due to the different $Q$s and $Z$s parameters.

Our prior distribution is therefore complemented with flat priors for those parameters that were not used in \citet{Klausner2025} (see Table~\ref{tab:prior} for a summary).
Incidentally, we note that the two surface parameters $G_0$ and $G_1$ and the spin-orbit parameter $w0$, while not useful for the description of nuclear matter, will explicitly enter in the description of nuclear clusters in the inner crust (see section~\ref{sec:crust}).

\begin{table}
\centering
\caption{Prior distribution for the parameters. 
         We marked with a * those parameters whose prior comes from the posterior distribution of \cite{Klausner2025}.
         For the others, we give the limits of this interval are shown.}
\label{tab:prior}
\begin{tabular}{llc}
\hline
$E_{sat}$ & $[$MeV$]$&*\\ 
$n_{sat}$ & $[$fm$^{-3}]$&*\\ 
$K_{sat}$ & $[$MeV$]$&  *\\ 
$Q_{sat}$ & $[$MeV$]$& [-2000,2000]\\ 
$Z_{sat}$ & $[$MeV$]$& [-3000,3000]\\ 
$E_{sym}$ & $[$MeV$]$&  * \\ 
$L_{sym}$ & $[$MeV$]$&  * \\ 
$Q_{sym}$ & $[$MeV$]$& [-4000,4000]\\ 
$Z_{sym}$ & $[$MeV$]$& [-5000,5000]\\ 
$m^*_{IS}$ & $[$-$]$&   *\\ 
$m^*_{IV}$ & $[$-$]$&   *\\ 
\hline
$w0$ & [MeV fm$^{5}$]& *\\ 
$G_0$ & [MeV fm$^{5}$]& *\\ 
$G_1$ & [MeV fm$^{5}$]& *\\ 
\hline
\end{tabular}
\end{table}

We extract 10$^5$ models from the prior distribution, assigning to each Skyrme a single random set of $Q_{sat,sym}$ and $Z_{sat,sym}$, and we compute the relative neutron star EoS.
We discard all the models for which the computation is unsuccessful (i.e., stability and causality are not respected~\cite{Montefusco2025}).   
We then assign to each successful model a likelihood  $\lk_{tot}$, which is the product of four independent likelihoods, $\lk_{tot}=\prod_{i}\lk_i$.
The $\lk_i$s encode the information from different independent physical constraints:
\begin{enumerate}[label=(\roman*)]
    \item $\lk_\text{J0348}$ is informed by the largest NS masses measured with high precision; we use that of pulsar PSR J0348+0432 from~\cite{Antoniadis2013};
    \item $\lk_\text{LVC}$ concerns the Ligo-Virgo collaboration tidal deformability measurement from the GW170817 event~\cite{Abbott2019};
    \item $\lk_\text{NICER}$ includes the joint inferences of the mass and radius of a NS by the NICER mission~\cite{Vinciguerra2024, Miller2021,Choudhury2024,Mauviard2025};
    \item $\lk_\chi$ measures the consistency between the energy per nucleon of neutron matter of the sampled model with the $\chi$-EFT chiral band presented in \cite{Huth2021} from a compilation of different ab-initio calculations (this $\chi$-filter is represented as a yellow region in Figure~\ref{fig:mapping}).
\end{enumerate}
The detailed mathematical description of these likelihoods is given in \cite{Montefusco2025}, see appendices A and B therein. The distribution of successful models, weighted with the total likelihood $\lk_{tot}$, will be our final result, the posterior distribution.

\section{Extended Thomas-Fermi approach for the inner crust}
\label{sec:crust}

The inner crust of an NS is thought to consist of a lattice of nuclei immersed in a gas of neutrons and free electrons~\cite{Chamel2008LRR, NS1book}.  
For each given value of the crustal baryonic density (or pressure), the composition is determined by the atomic number of the nuclei and the densities of the surrounding electron and neutron gases.

Since the seminal work of \citet{Negele1973}, studies have focused on finding the optimal energy configuration of the unit cell - the so-called Wigner-Seitz (WS) cell.
Due to the difficulties in calculating the inhomogeneous crust, EoSs employed for astrophysical inference are often matched ~\cite{Greif2019, Essick2021, Raithel2023, Huang2024} to a unique given crust, typically the Liquid Drop Model (LDM) based crust calculation of~\cite{Baym1971, Douchin01}.
However, a unified EoS employing the same nuclear functional in the crust and in the (outer) core is necessary to have statistically meaningful predictions of crustal properties, as well as precise estimations of NS radii~\cite{Fortin2016,Davis2024}. 

To our knowledge, all previous Bayesian studies of NS crustal properties using unified EoS's \cite{Carreau2019,Lim2019,DinhThi2021,Newton2022} employed the simple Compressible Liquid Drop Model (CLDM) description of the WS cell as pioneered in~\cite{Baym1971}.
As explained in Section~\ref{sec:bayes}, our nuclear models are originally Skyrme models fitted to nuclear data: we can, therefore, approach the problem more microscopically.
To this end, we compute the baryonic contribution $E_{WS}^{nuc}$ to the total WS energy employing the full Skyrme energy density functional approximated in extended Thomas-Fermi (ETF)~\cite{Brack1985,Pi1986,Onsi1997,Grill2014,Aymard2014}:
\begin{equation}
    E_{WS}^{nuc}=4 \pi \int^{R_{WS}}_0 \!\!\! r^2 \,
    \mathcal{E}^{Sky}_{ETF}
    \left( n(r), n_p(r) \right)  dr \, ,
   \label{eq:ws_energy}
\end{equation}
where $\mathcal{E}^{Sky}_{ETF}$ is the Skyrme energy density functional, and the subscript ETF indicates that the non-local operators (kinetic energy and currents) are expanded to second order in $\hbar$, thus becoming simple functionals of the local densities $n(r)$ and $ n_p(r)$.
In the above equation, $R_{WS}$ is the radius of the spherical WS cell, as detailed in Appendix~\ref{app:crust}.
We stress that the surface and spin-orbit parameters explicitly enter into $\mathcal{E}_{Sky}^{ETF}$. 
Their prior distribution comes from the nuclear structure data inference: all the information obtained in \cite{Klausner2025} is carried over into the crust construction.

The two density profiles $n(r)$ and $n_p(r)$ in \eqref{eq:ws_energy} are parametrized using Woods-Saxon functions~\cite{Aymard2014}: minimizing the cell energy with respect to the density profile parameters at fixed baryon density $n_B$ yields the composition and EoS of the inner crust of the star.  
In practice, this involves solving a system of nonlinear equations to find the stationary point of $E_{WS}^{nuc}$, see Appendix~\ref{app:crust} for details.  
We have verified that, when applied to the Sly4 Skyrme interaction~\cite{Chabanat1997}, our method produces results in good agreement with those obtained from the more sophisticated ETF code of~\cite{Pearson2018, Shchechilin2024} (see also Appendix~\ref{app:crust}).

To estimate the crust-core (CC) transition point, we compare at each density point the energy density of the optimal ETF solution for the clusterized phase $\varepsilon_{WS}$ with the energy density of the homogenous phase $\varepsilon_{core}$.
The condition $\Delta \varepsilon = \varepsilon_{WS} - \varepsilon_{core} > 0$ marks the transition to the core.

Unfortunately, in the random generation of the EoS model for the Bayesian inference, the algorithm often encounters numerical instabilities and cannot find a solution to the system of equations at densities of about $0.06$ fm$^{-3}$, where we still have $\Delta \varepsilon < 0$.
It is possible to mitigate this problem by tuning the initial guesses for the minimisation procedure, refining the mesh in baryon densities, or implementing ad hoc back-stepping conditions.
However, we have not been able to devise a model-independent solution that can be implemented in an automated procedure for our Bayesian study, namely a robust minimization procedure that works for all baryon densities and all instances of nuclear models we need to sample.
We instead resorted to a linear extrapolation of the quantity $\Delta\varepsilon (n_B)$ to estimate the value of the CC transition density~$\ncc$. 
The validity and limitations of this extrapolation are further discussed below and in Section~\ref{subsection_CC}.
\begin{figure}
    \includegraphics[width=0.99\linewidth]{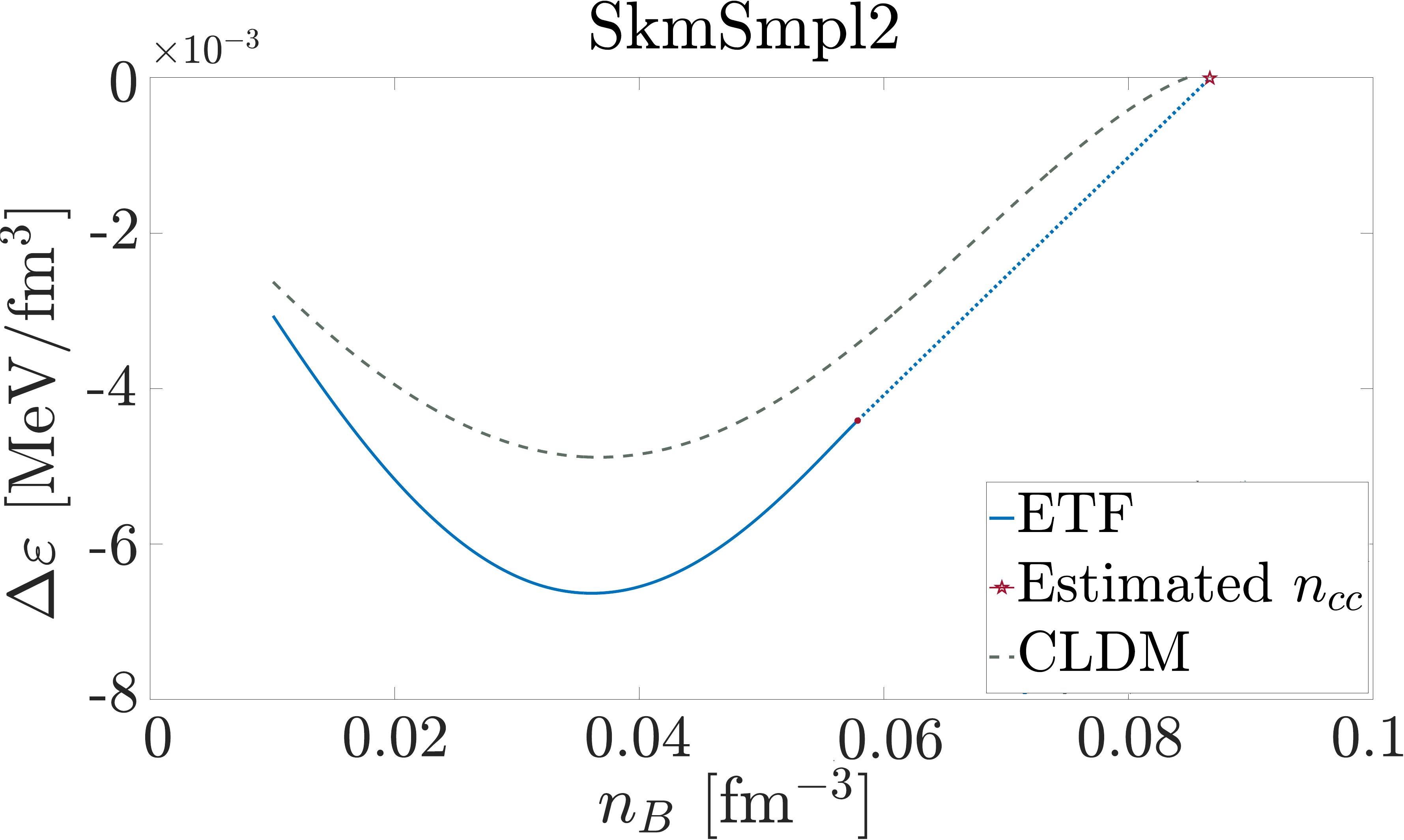}
    \caption{Difference in energy density $\Delta\varepsilon$ between the homogenous and clusterized phase, as a function of the baryon density $n_B$, for the sampled Skyrme interactions SkmSmpl2. 
    The blue line is the ETF computation, while the grey dashed one is its CLDM counterpart.
    The red dot marks when the ETF computations fail; the linearly guessed $\ncc$ is marked with a red star.}
    \label{fig:lienar_extrapolation}
\end{figure}
Once  $\ncc$ is determined, we can interpolate the energy density $\varepsilon$, the pressure $P$, the proton fraction $y_p$, and the free neutron gas density $n_g$.
The first two are interpolated with a third-order polynomial, ensuring the continuity of $\varepsilon$ and $P$. 
On the other hand, $y_p$ shows an almost linear relation with the baryon density, and so we interpolated it linearly.
Finally, $n_g$, which at $\ncc$ coincides with the total neutron density, is interpolated between the last five points of the crust and the first five of the core with a third-order polynomial.

The extrapolation procedure is illustrated in Figure~\ref{fig:lienar_extrapolation} for a representative model example (SkmSmpl2; see Appendix~\ref{app:sampled_skyrme} for details).  
In the figure, the blue line shows the energy density difference $\Delta\varepsilon$ as a function of the baryon density $n_B$ from the ETF calculation: in this typical case, the computation stops slightly above 0.06~fm$^{-3}$ (red point), while the dotted line represents the extrapolation, and the red star marks the estimated $\ncc$.  
For comparison, the dashed grey line shows $\Delta\varepsilon$ computed using the CLDM\footnote{
    All CLDM calculations in this work are performed using the code developed in~\cite{DinhThi2021b}, to which we refer for implementation details.
}.  
The ETF curve closely resembles the CLDM one in shape, and the trend appears approximately linear in the region where the ETF calculation breaks down, which supports the validity of the extrapolation procedure.

\begin{table}
\centering
\caption{Crust-Core transition for SLy4 interaction with different computational strategies.}
\label{tab:sly4_ncc}
\begin{tabular}{lcc}
\hline
 Strategy & Source    &  $\ncc$ [fm$^{-3}$]    \\   
\hline
CLDM & \cite{Carreau2019} & 0.073-0.083   \\ 
CLDM & \cite{Vinas2017} & 0.072   \\ 
CLDM & \cite{Douchin01} & 0.076   \\ 
CLDM & \cite{DinhThi2021b} & 0.081   \\
\hline
Spinodal (thermodynamical) & \cite{Ducoin2011} &0.089   \\ 
Spinodal (dynamical) & \cite{Ducoin2011}& 0.080 \\ 
\hline
Full ETF & \cite{Martin2015}&0.081 \\
Full ETF & \cite{Shchechilin2024,Shchechilin2025} & 0.082
\\ 
ETF & this work & 0.083\\ 
\hline
\end{tabular}
\end{table}

Another validation test is presented in Table~\ref{tab:sly4_ncc}, which reports different $\ncc$ calculations for the widely employed Skyrme functional SLy4 \cite{Chabanat1997}, and compares them with our estimated value.
We can see that the linear extrapolation leads to an agreement with the full ETF result of \cite{Martin2015} within $\approx 2.5\%$. 
We can also see that the CLDM predictions are affected by a considerable dispersion, due to the different possible surface energy representations in that classical approach.
Even within this dispersion, in the case of the Sly4 model, the more microscopically funded ETF formalism tends to produce slightly higher values for the CC transition point. The dependence on the nuclear model will be studied in detail in Section \ref{subsection_CC}.

Table~\ref{tab:sly4_ncc} also reports the CC transition density obtained with two other methods widely employed in the literature, the thermodynamical and dynamical spinodal instability boundaries~\cite{Ducoin2011}:
\begin{enumerate}
    \item[-] The thermodynamical spinodal corresponds to the instability point of neutral nuclear matter undergoing the nuclear liquid-gas phase transition. 
    It provides an upper bound to the physical instability of homogeneous $npe\mu$ matter to cluster formation, due to the absence of finite-size (surface and Coulomb) effects~\cite{Pethick1995}. 
    These effects are included in the dynamical spinodal calculation, which determines when matter becomes unstable against finite-size density fluctuations, that is, against the clusterization of matter in the presence of a uniform electron background~\cite{Ducoin2007,Ducoin2011}.
    \item[-] The dynamical spinodal can be considered a lower limit to the CC transition point, since in a first-order phase transition the unstable region is systematically contained within the coexistence region, and the crust formation timescale is expected to be sufficiently long for metastable homogeneous fluid matter to evolve into solid crustal matter~\cite{haensel_book}. 
\end{enumerate}
As seen in Table~\ref{tab:sly4_ncc}, in the case of Sly4, our extrapolated ETF result aligns with these qualitative considerations.
Among the various methods used in the literature to define the crustal EoS and the CC transition point, explicit crust modelling with microscopic many-body approaches (HFB~\cite{Pastore2017} or its semi-classical counterpart, ETF) is considered the most accurate.
However, concerning the determination of the CC point, we cannot exclude a possible bias introduced by the linear extrapolation of our ETF results. 
For this reason, we adopt the ETF method to compute the crust and its EoS, but also consider the CLDM and the spinodal definition in determining the crustal properties in Section~\ref{sec:res}.

\section{Results}
\label{sec:res}

In this section, we present the results of our Bayesian analysis for NS properties, including the information of the nuclear structure experiments (that correspond to our prior distributions), from the ab-initio calculations of neutron matter with chiral interaction, and from the astrophysical observations.  

\subsection{Nuclear matter parameters}

\begin{figure*}
    \centering
    \subfloat[Isoscalar parameters]{
    \includegraphics[width=.4635\linewidth]{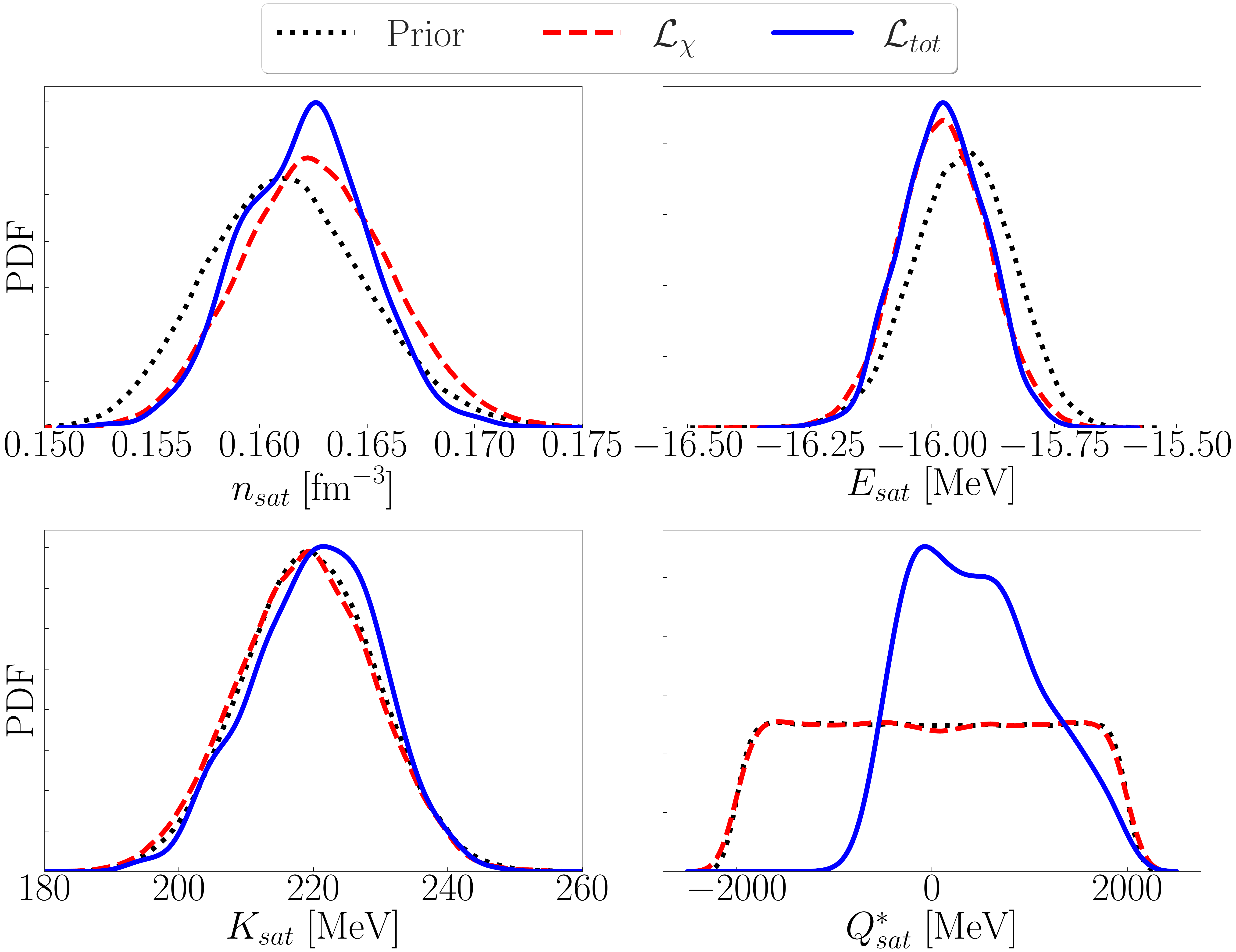}}
    \qquad
    \subfloat[Isovector parameters]{
    \includegraphics[width=.4565\linewidth]{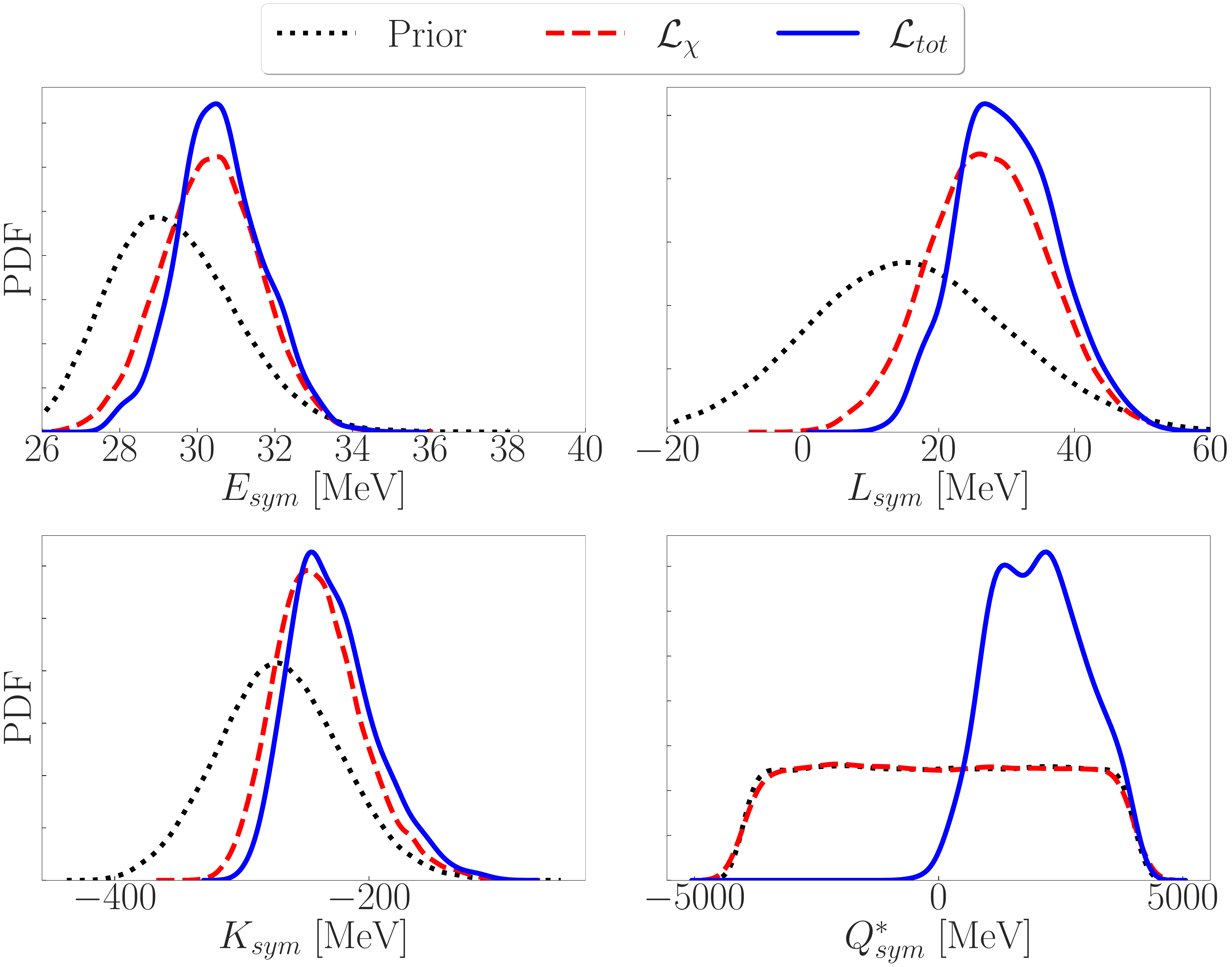}}
    \\
    \subfloat[Surface parameters
    \qquad\qquad\qquad\qquad\qquad\qquad\qquad\qquad\qquad\qquad\qquad
    (d) Effective mass parameters]{
    \includegraphics[width=.97\linewidth]{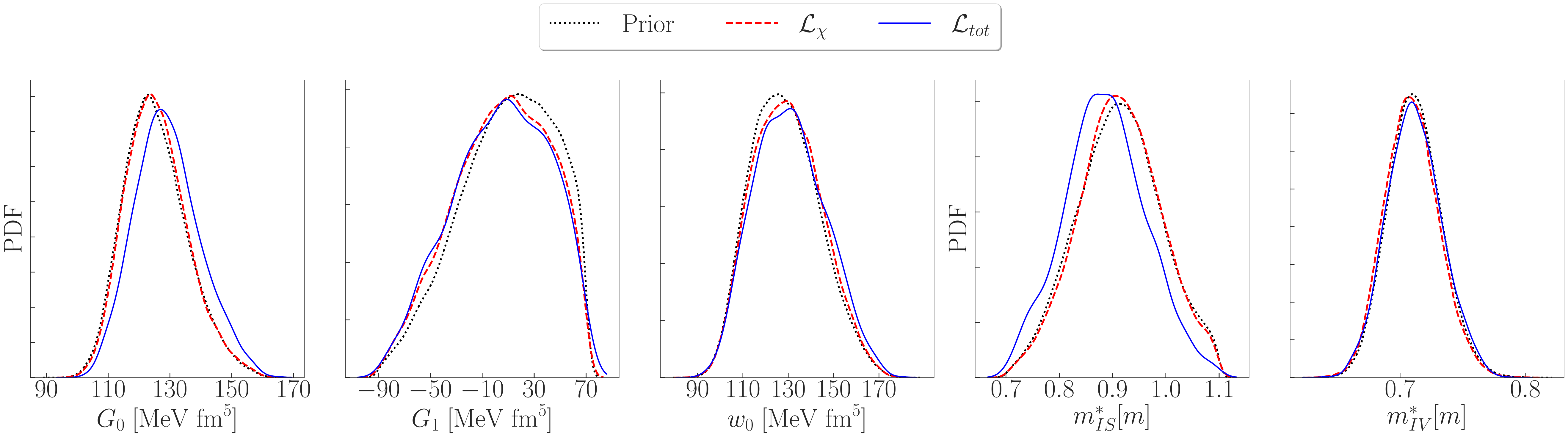}}
    \caption{Nuclear matter parameters marginalized priors and posteriors. Bulk isoscalar (a), bulk isovector (b), surface and spin-orbit (c), and effective masses (d) parameters are shown.}
    \label{fig:marg_posteriors}
\end{figure*}

In Figure~\ref{fig:marg_posteriors}, we show the marginalized posterior distributions of the different parameters of the Skyrme-metamodel. 
The current constraints bear no information on the fourth-order parameters $Z_{sat,sym}$, so we omitted them in the figure. 
The dotted black lines correspond to the marginalized prior informed by nuclear experiments, while the dotted red ones represent the prior weighted only with $\mathcal{L}_\chi$, which is additionally informed by the chiral calculations; finally, the solid blue line shows the marginalized full posteriors that also includes the astrophysical information.

Looking at the isoscalar parameters in the upper left part, we see that prior and posterior distributions are very similar, except for $\qsat$, which is shifted by the astrophysical constraints toward the positive side of its agnostic prior distribution. 
This indicates that the isoscalar sector is already well constrained by nuclear physics data, except at high-density, where the requirement to support the gravitational pressure of very massive NSs leads to a stiffening of the EoS, and consequently to positive values for both $Q_{sat}$ and $Q_{sym}$\footnote{The $Q_{sat,sym}$ parameters displayed are the ones defined above saturation.}.

The complementarity of the information provided by nuclear theory and nuclear experiments is clearly visible in the low-order isovector bulk parameters (upper right panels of Figure~\ref{fig:marg_posteriors}).
While the information from nuclear experiments is compatible with the chiral constraint, the latter leads to a sharper definition of the parameters and an average shift toward higher values. 
The peak of the $E_{sym}$ distribution increases from 29 MeV to almost 31 MeV, while the $L_{sym}$ most probable value shifts from $\ approx15$ MeV to $\ approx30$ MeV. 
Interestingly, the application of the astrophysical constraints only induces very minor modifications to these distributions, highlighting the strong predictive power of nuclear theory for extremely neutron-rich matter.
The posterior distribution of $\ksym$ is also shifted to higher values compared to the nuclear experiments informed prior, but little can be concluded from this, since $\ksym$ is not an independent parameter in Skyrme interactions (see the discussion in Section~\ref{sec:bayes}).

The lower part of Figure~\ref{fig:marg_posteriors} shows the isoscalar and isovector surface parameters of the Skyrme interaction, $G_0$ and $G_1$, the spin-orbit coupling $w_0$, and the two effective masses $m^*_{IS,IV}$. 
These parameters are already well constrained by the nuclear structure inference of~\cite{Klausner2025} (``prior'' curves in Figure~\ref{fig:marg_posteriors}), and we observe that none of the additional constraints modifies their distributions.
This indicates that the requirement of supporting the crust does not impose further constraints on the nuclear models.

\begin{figure*}
    \subfloat[]{
    \includegraphics[width=.6\linewidth]{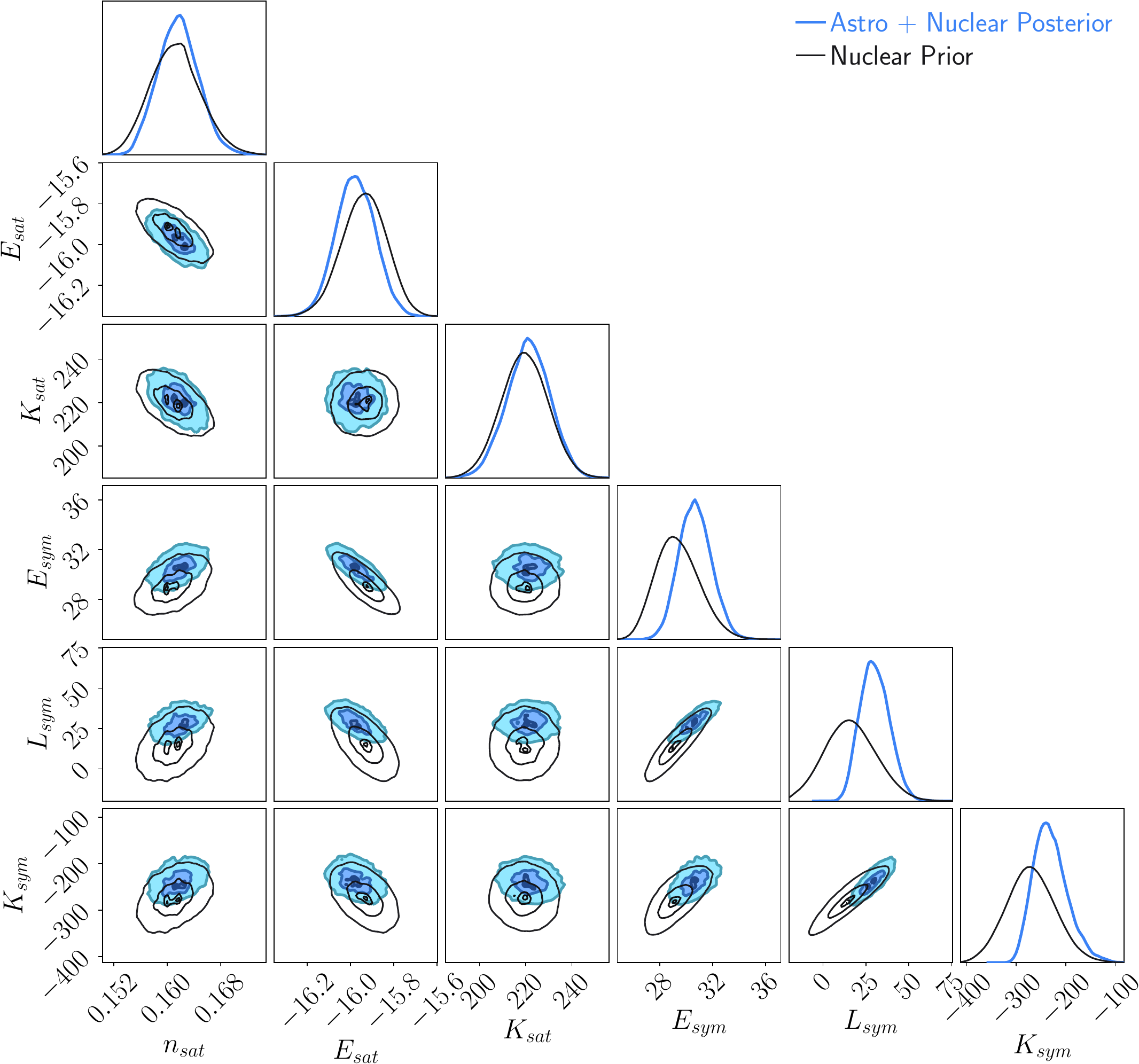}}
    \\
    \subfloat[]{
    \includegraphics[width=.6\linewidth]{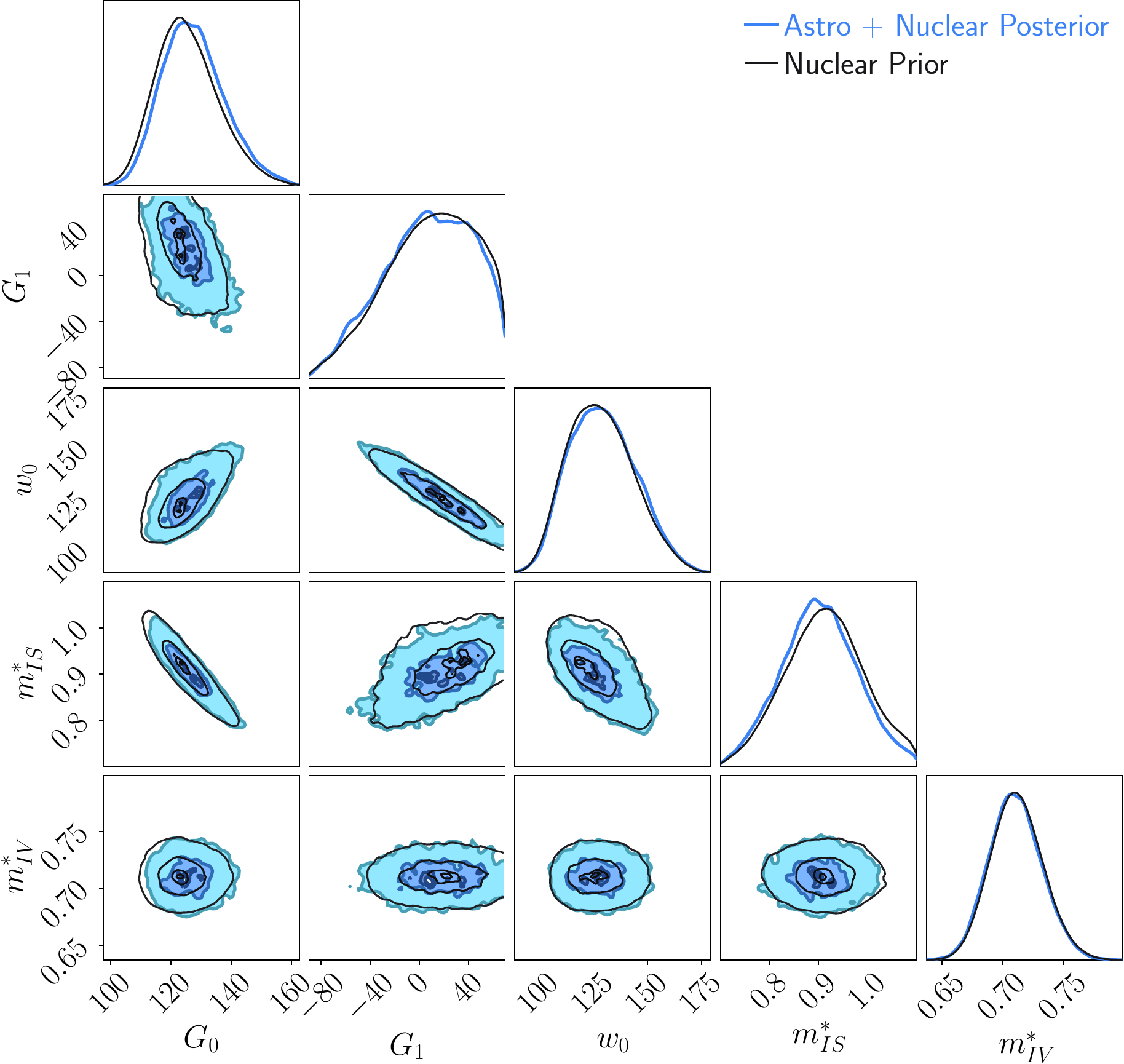}}
    \caption{
    Corner plots of the bulk (a) and surface and effective mass (b) parameters. Black lines and contours: prior. Blue lines and contours: posterior. 
    Other possible correlations were not significant and are therefore not shown.
    }
    \label{fig:corner_plots}
\end{figure*}

In Figure~\ref{fig:corner_plots}, we show selected corner plots illustrating the most relevant correlations among the bulk and surface model parameters. The additional information preserves all correlations imposed in the prior by experimental nuclear observables from ab-initio calculations and observational constraints, but the parameter estimation is significantly narrowed in the isovector sector.

\subsection{Crust-Core transition point}
\label{subsection_CC}

\begin{figure*}
    \subfloat[ETF vs. CLDM]{
    \includegraphics[width=.55\linewidth]{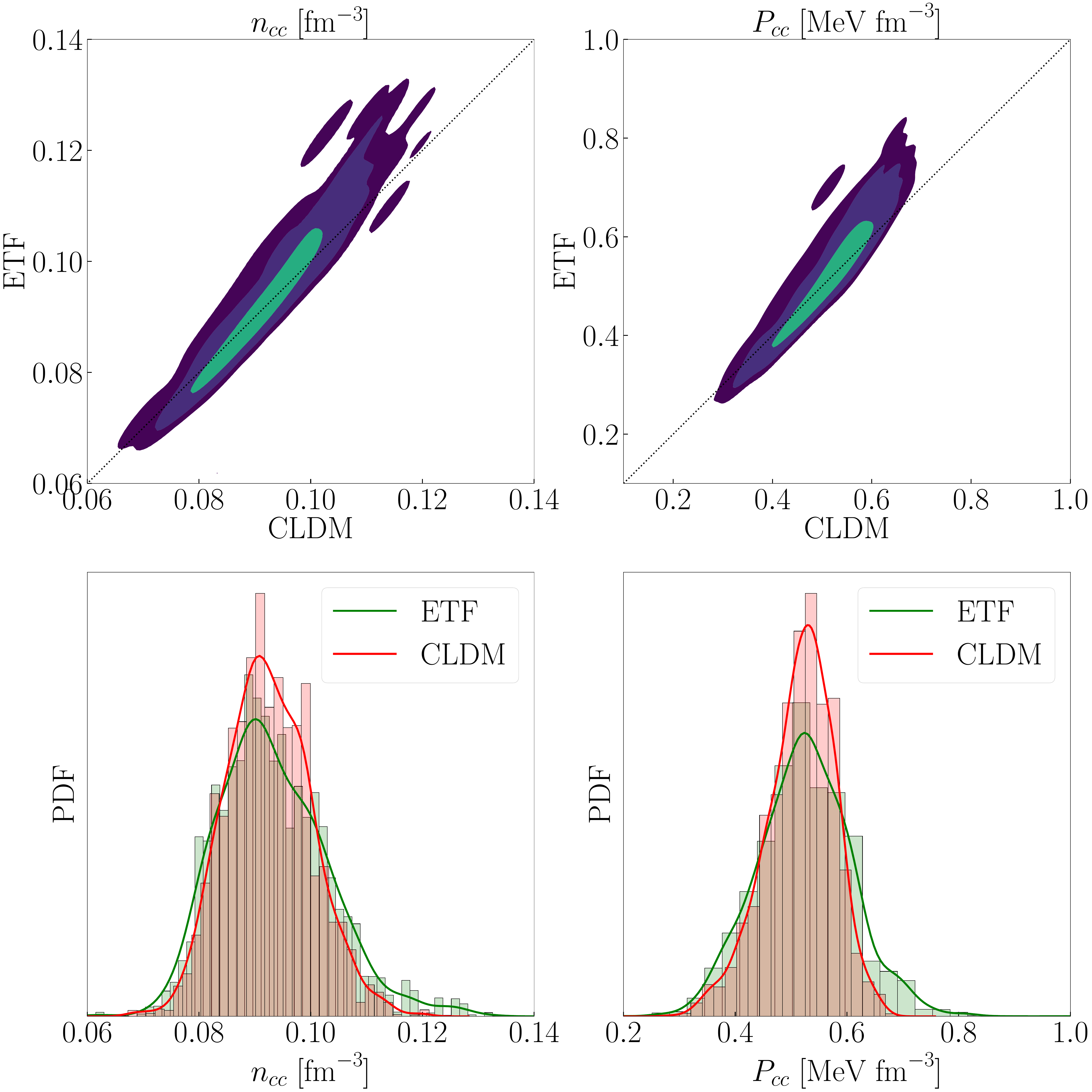}}  
    \\
    \subfloat[ETF vs. Spinodal]{
    \includegraphics[width=.55\linewidth]{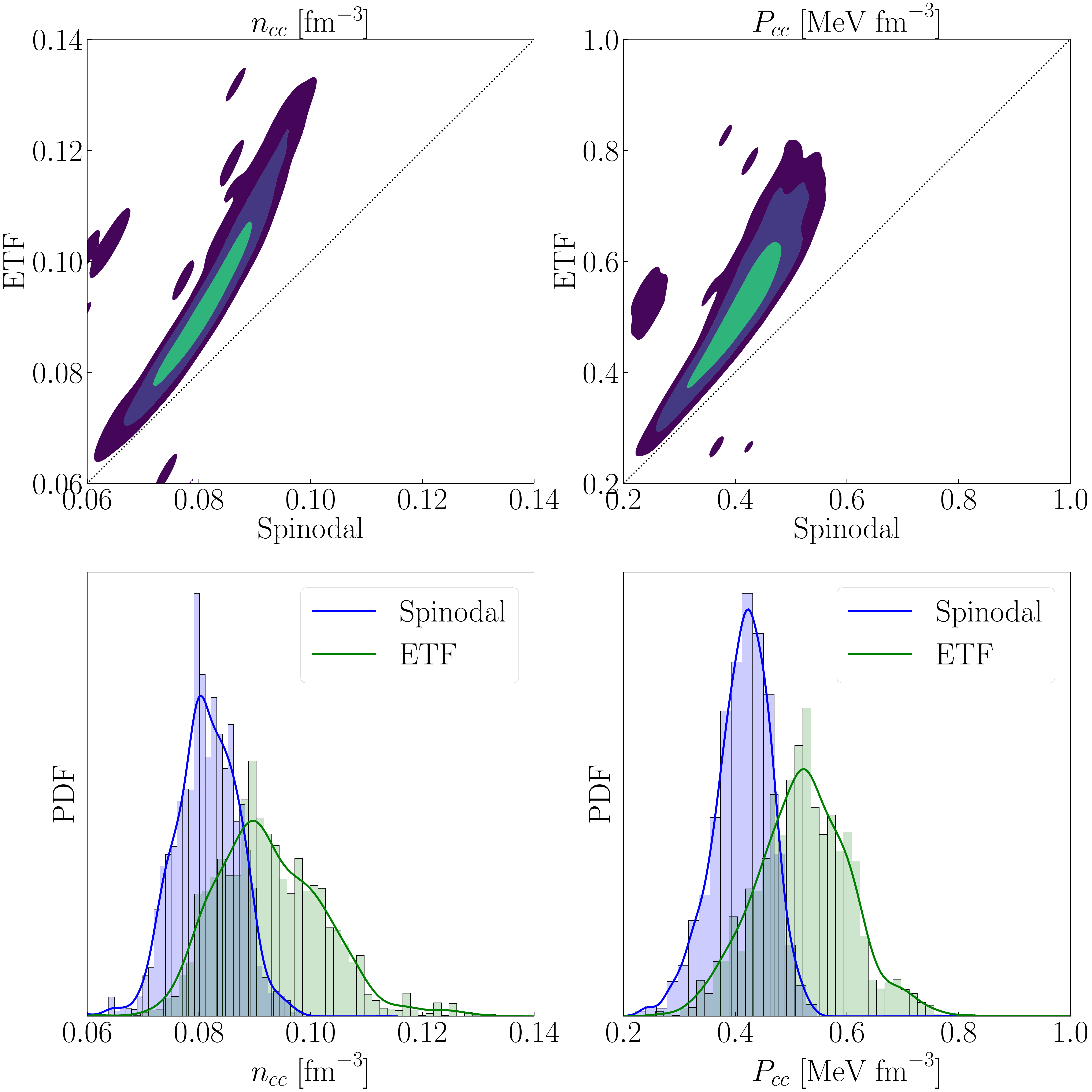}}
    \caption{Posterior of the CC density and pressure transition point obtained with different estimation methods. Left panels: comparison between ETF (with linear extrapolation, see text) and CLDM. Right panels: comparison with the dynamical spinodal definition.}  
    \label{fig:crust-core_transition}
\end{figure*}

The Bayesian posterior for the CC transition point is shown in Figure~\ref{fig:crust-core_transition}, where we show the joint probability distribution of the CC transition density $\ncc$ and the corresponding pressure $P_{cc}$, along with the marginalized distributions obtained using the different methods (ETF, CLDM, and dynamical spinodal) described in Section~\ref{sec:crust}.
In Table \ref{tab:ncc_Pcc_comparison_table} we report the median value and 68\% confidence interval of the distributions.
\begin{table}
\centering
\caption{Median value and 68\% confidence interval limits of the crust-core transition density $n_{cc}$ and the pressure at transition $P_{cc}$, for the three different estimation methods.}
\label{tab:ncc_Pcc_comparison_table}
\begin{tabular}{llll}
\hline
	 & ETF & CLDM & Spinodal \\ 
\hline
$n_{cc}$ [fm$^{-3}$] & $0.092^{+0.010}_{-0.009}$ & $0.095^{+0.009}_{-0.008}$ & $0.082^{+0.006}_{-0.006}$  \\ 
$P_{cc}$ [MeV fm$^{-3}$] & $0.525^{+0.079}_{-0.083}$ & $0.533^{+0.054}_{-0.067}$ & $0.418^{+0.044}_{-0.054}$  \\ 
\hline
\end{tabular}
\end{table}

The ETF estimation exhibits a tail at high transition densities, $\ncc \gtrapprox 0.11$~fm$^{-3}$, as well as some outliers corresponding to values of $\ncc$ significantly higher than those from CLDM or the spinodal. 
This is a spurious effect caused by the linear extrapolation used in the ETF to determine the transition point. 
This interpretation is justified by the fact that, for a minority of sampled models, the numerical solution of the ETF variational equations fails near the minimum of $\Delta\varepsilon$ (see Figure~\ref{fig:lienar_extrapolation}), leading the extrapolation to overestimate the transition density, sometimes substantially.

Apart from this bias, the distributions obtained with CLDM and ETF, and shown in the left part of Figure ~\ref{fig:crust-core_transition} are very similar. 
It might seem surprising that a simple approach like the CLDM yields results in reasonable agreement with ETF. 
However, it is important to recall that: 
\begin{itemize}
    \item[$-$] The CLDM is informed by the same bulk parameter distributions as the ETF. 
    \item[$-$] Some level of consistency between surface and bulk parameters is ensured in our version of the CLDM, since its surface parameters are optimized on experimental nuclear masses~\cite{DinhThi2021b}.
    \item[$-$] The CLDM method is affected by important uncertainties due to the different possible modelings of the surface energy (see Table \ref{tab:sly4_ncc}).
    However, in the version used in the present paper and taken from ref.~\cite{DinhThi2021b}, the parameter $p$ governing the behavior of the surface energy at extreme isospin ratios is fixed to $p=3$, a value suggested in the seminal CLDM papers~\cite{Ravenhall_1983,Lattimer_1991} to better comply with Thomas-Fermi calculations in slab geometry.  
\end{itemize}
The right part of Figure~\ref{fig:crust-core_transition} compares the ETF estimation of the CC transition point with that obtained from the dynamical spinodal instability~\cite{Ducoin2007}.
Before comparing the CC transition obtained with the spinodal and ETF methods, we first comment on our results for the dynamical spinodal. 
A previous Bayesian study based on the spinodal criterion~\cite{Antic2019} reported significantly lower values, $\ncc = (0.071 \pm 0.011)$~fm$^{-3}$ and $P_{CC} = (0.294 \pm 0.102)$~MeV~fm$^{-3}$. 
That study incorporated the chiral constraint, but the surface parameters were constrained solely by a global fit to experimental masses using a classical approach. 
As a result, the posterior values reported in~\cite{Antic2019}, $G_0 = 30 \pm 9$ and $G_1 = 115 \pm 143$, are in tension with the more robust estimates of~\cite{Klausner2025}, namely $G_0 = 125 \pm 10$ and $G_1 = 9 \pm 36$. 
This discrepancy underscores the importance of accurately determining the surface parameters when predicting the CC transition point.
 
Turning to the comparison between spinodal and ETF results, we observe that even with optimized surface parameters, the spinodal estimate underestimates both the dispersion and the mean values of the $\ncc$ and $P_{cc}$ posteriors. 
This is in qualitative agreement with the expectation that the spinodal criterion provides a lower bound to the transition point, as discussed in Section~\ref{sec:crust}. 
We attribute the observed discrepancy to intrinsic physical differences rather than to the extrapolation of $\ncc$. 
The CLDM distributions of $\ncc$ and $P_{cc}$ are much closer to those of the ETF than to those of the spinodal instability.
These findings highlight the need for a unified treatment of the EoS and explicit modelling of the inhomogeneous crust to determine the CC transition and related crustal properties reliably. 

\subsection{Crust properties} 

\begin{figure}
    \subfloat[ Cluster mass number $A$ (upper left), atomic number $Z$ (upper right), central density $n_0$ (middle left), gas density $n_g$ (middle right), and Wigner-Seitz cell radius $R_{WS}$ (lower panel).  ]{
    \includegraphics[width=.9\linewidth]{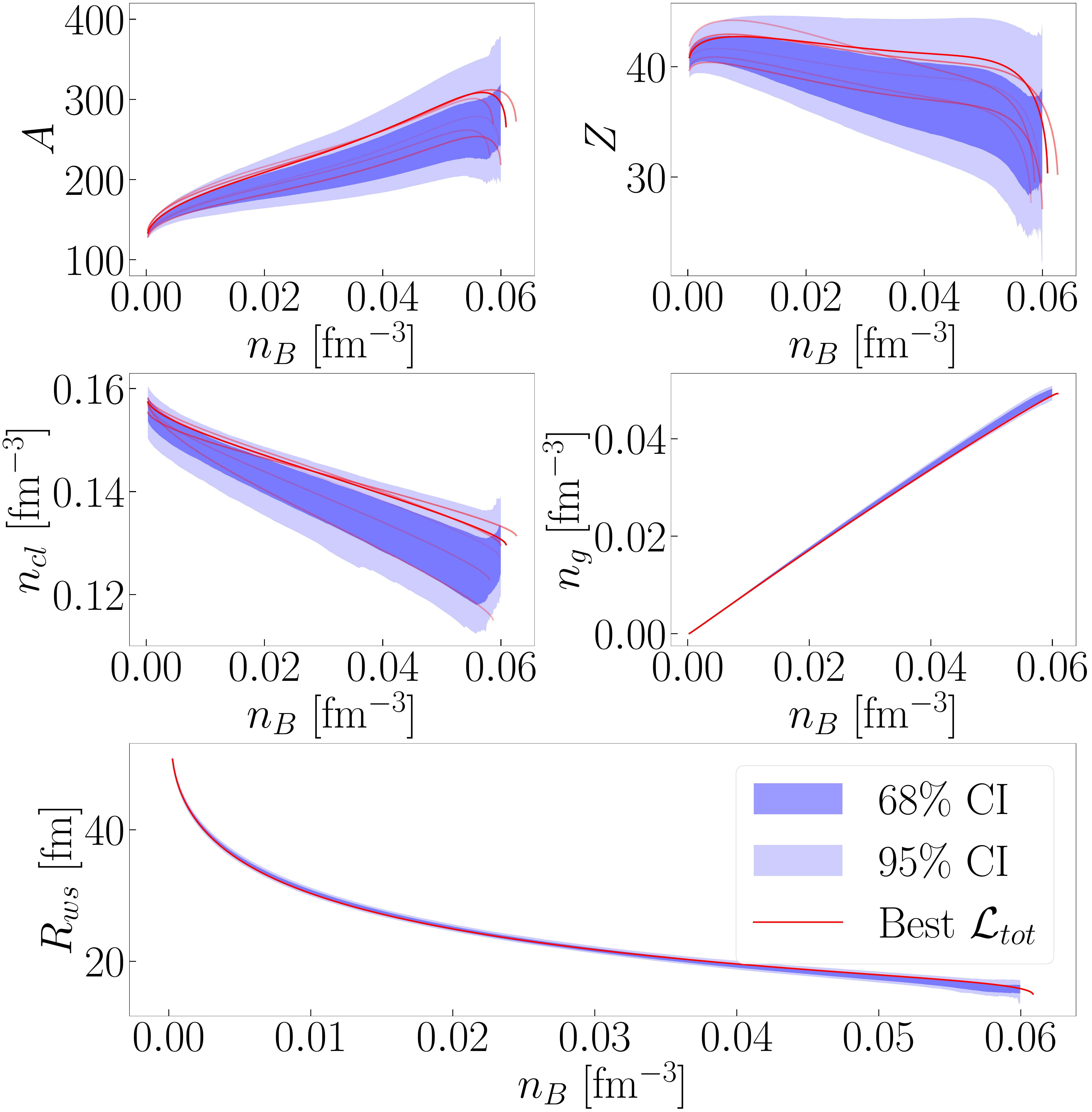}
    \label{fig:crust_composition_posterior}}
    \\
    \subfloat[  Volume fraction occupied by the cluster in the cell (upper left), fraction of nucleons  (upper right), and protons  (lower panel). ]{
    \includegraphics[width=.9\linewidth]{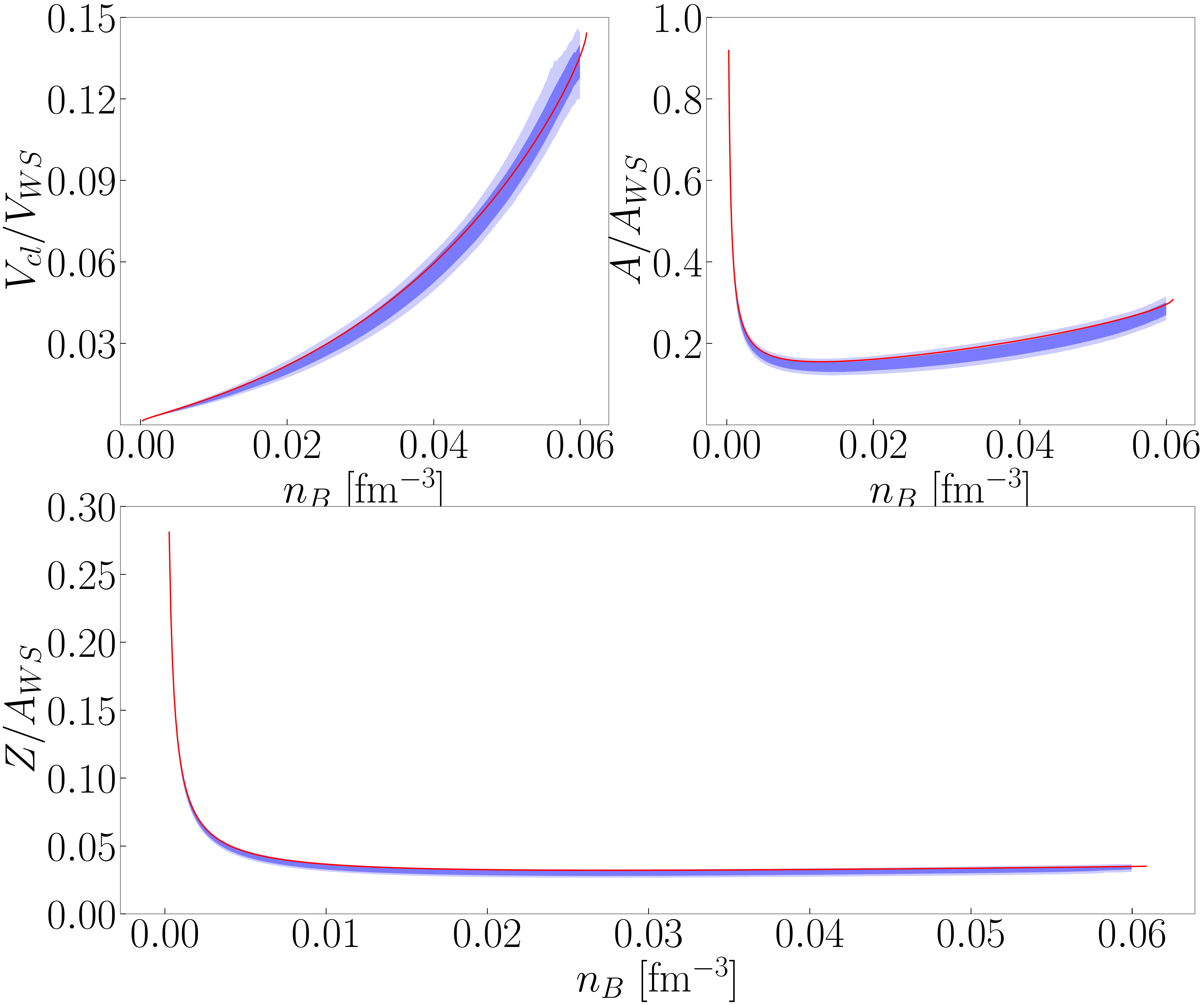}
    \label{fig:crust_composition_fractions}}
    \caption{ Posterior distribution of the crust composition. At each baryon density $n_B$, the blue bands indicate the 68\% and 95\% confidence intervals. The red lines correspond to models with the highest $\lk_{tot}$;  opacity increases with increasing~$\lk_{tot}$.}
    \label{fig:crust_composition}
\end{figure}

In Figure~\ref{fig:crust_composition_posterior}, we present our results for the crust composition, obtained using the ETF variational method. 
We focus on five quantities that are sufficient to characterize the crust composition (see Appendix~\ref{app:crust} for details): the number of nucleons in the cluster $A$, the number of protons $Z$, the nuclear density at the cluster center $n_0$, the neutron gas density $n_g$, and the radius of the Wigner-Seitz cell $R_{WS}$. 
The plots are truncated at $n_B = 0.06$~fm$^{-3}$, as the information at $n_B > 0.06$~fm$^{-3}$ is affected by a lack of statistics due to numerical issues in solving the system of variational equations.

It is interesting to observe that the best model lies at the edge of the 95\% region for both $A$ and $Z$. 
While this may seem counterintuitive, we recall that $\lk_{tot}$ is the product of four terms (see Section~\ref{sec:bayes}), none of which explicitly involves the crust. 
This explains why high-likelihood models are broadly distributed across the posterior and not clustered around their central region.

The crust composition's qualitative behavior as a density function is consistent with previous studies based on a more limited set of models and constraints~\cite{Carreau2019,Grams2021}.
The progressive melting of clusters toward the core is indicated by an increasing cluster mass number $A$ and a decreasing central density $n_0$, while the decrease in atomic number $Z$ reflects the increasing neutron richness at higher densities imposed by beta equilibrium. 

The specificity of our work lies in the ability to assess, from the Bayesian posterior, the residual model dependence of the crust composition after incorporating all available information from nuclear theory and experiment into the ETF formalism. 
In particular, we observe that the steadily increasing neutron gas density and the steadily decreasing linear size of the WS cell are quasi-universal features. 
Their predicted values closely match those obtained with the more sophisticated ETFSI formalism~\cite{Potekhin2013,Pearson2018}, which accounts for proton shell effects within the Strutinsky approach.

A much stronger model dependence is instead observed for the cluster properties $A$, $Z$, and $n_0$. 
This is not surprising, given that the number of protons (nucleons) in the cluster accounts for no more than $\approx 3\%$ ($\approx 30\%$) of the total baryon number in the Wigner-Seitz cell, as we can appreciate in Figure \ref{fig:crust_composition_fractions}.
This, combined with the nuclear physics uncertainties associated with neutron-rich nuclei, results in a high dispersion of the predictions. 
The smooth behavior of $Z$ reflects the semiclassical nature of the ETF approximation, which does not account for shell effects. 
In contrast, the more microscopic  ETFSI approach is known to stabilize specific $Z$ numbers corresponding to shell closures over broad regions of the inner crust (typically $Z=40$ ~\cite{Pearson2018}). 
Nevertheless, the dispersion arising from model dependence exceeds the impact of shell effects that are here neglected~\cite{Carreau2020}.

\begin{figure*}
    \subfloat[Radius]{
    \includegraphics[width=0.48\linewidth]{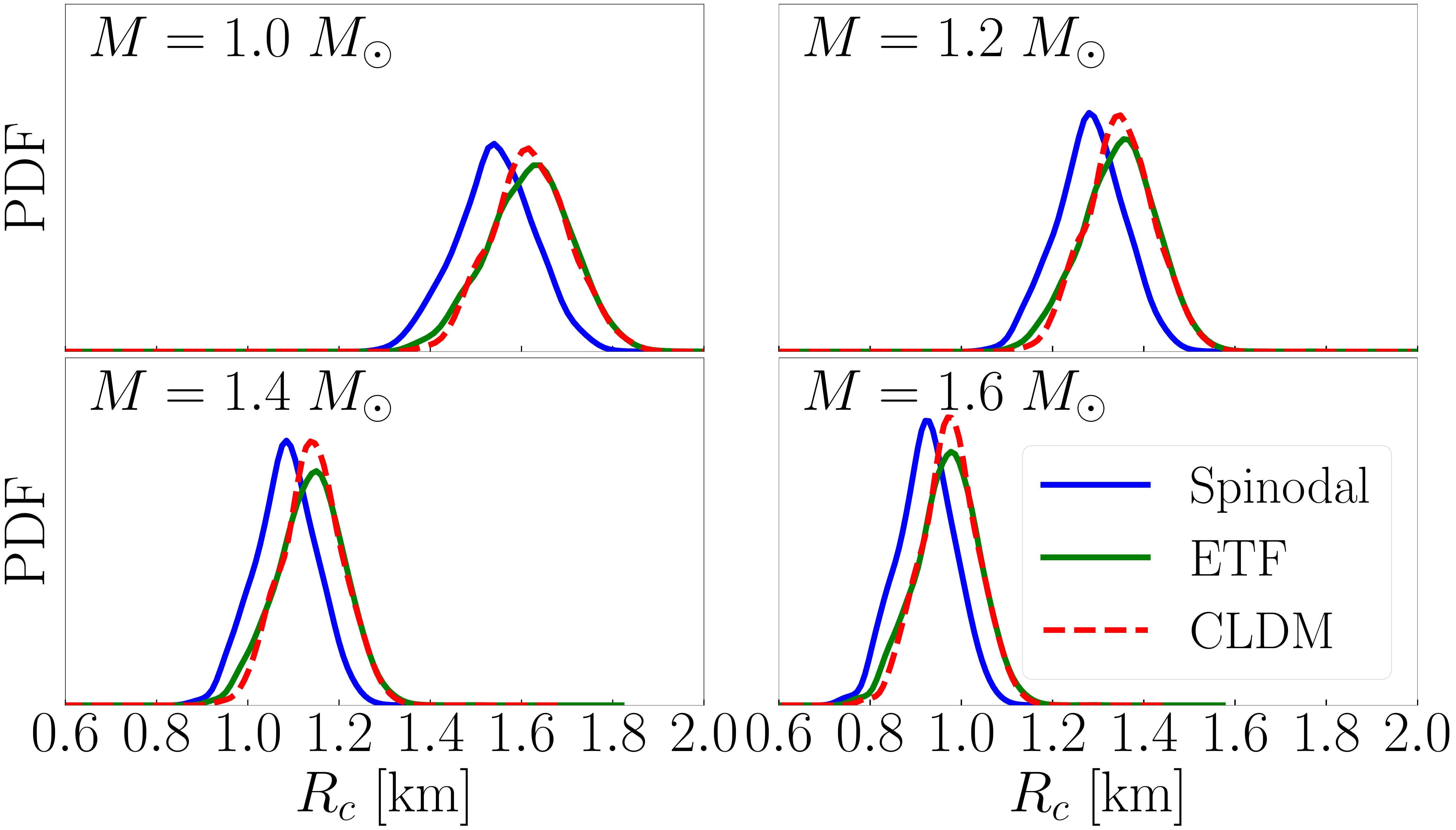}}
    \quad
    \subfloat[Crust Moment of inertia]{
    \includegraphics[width=.48\linewidth]{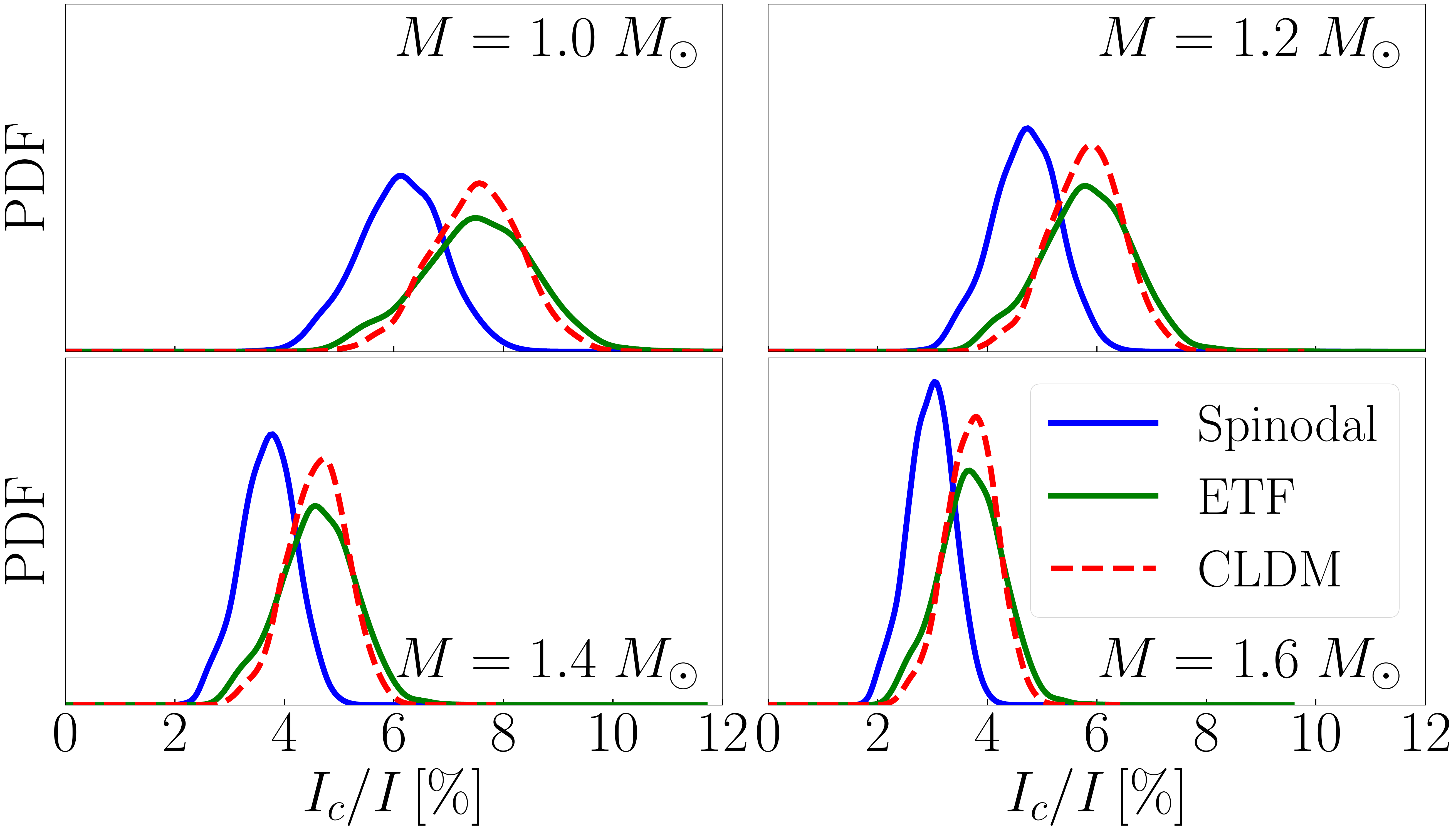}}
    \caption{Posterior distributions of the crustal radius (left) and moment of inertia (right) for four different NS masses, using different definitions of the CC transition point: ETF (green), dynamical spinodal (blue), and the CLDM~(dashed~red).}
    \label{fig:crust_I_r}
\end{figure*}

From an astrophysical perspective, the quantities of greatest interest are the crustal radius $R_c$ and the crustal moment of inertia $I_c$, which are key parameters in simulations of NS cooling~\cite{Page2013}. 
The posterior distributions of these quantities are shown in Figure~\ref{fig:crust_I_r}, based on the different definitions of the CC transition point introduced in Section~\ref{sec:crust} and for various NS masses. 
We focus in particular on relatively low masses, where the crust plays a more significant role.

We observe that the radius predictions from ETF and CLDM are very close, consistent with the compatible CC transition point values obtained in the two approaches, as shown in Figure~\ref{fig:crust-core_transition}. 
Conversely, the spinodal definition systematically yields lower values for both $R_c$ and $I_c$, again due to the shift in the transition point.

Finally, the differences between the methods tend to diminish as the stellar mass increases, due to the reduced importance of the crust. 
The median value and 68\% confidence interval of the distributions are reported in Table \ref{tab:moments_and_radii_non_piff}.
\begin{table}
\centering
\caption{Median value and 68\% confidence interval limits of the crust radius $R_c$ and the fraction of the moment of inertia of the crust over the total $I_c/I$, for four different NS masses.}
\label{tab:moments_and_radii_non_piff}
\begin{tabular}{ccccc}
\hline
	 &  $M$ $[M\odot]$ & ETF & CLDM & Spinodal \\ 
\hline
\multirow{4}{*}{$R_c$ [km]} & 1.0 & $1.622^{+0.087}_{-0.095}$ & $1.619^{+0.085}_{-0.088}$ & $1.542^{+0.086}_{-0.084}$  \\ 
  & 1.2 & $1.349^{+0.077}_{-0.083}$ & $1.349^{+0.074}_{-0.078}$ & $1.282^{+0.076}_{-0.076}$  \\ 
  & 1.4 & $1.141^{+0.068}_{-0.079}$ & $1.141^{+0.067}_{-0.071}$ & $1.084^{+0.067}_{-0.069}$  \\ 
  & 1.6 & $0.973^{+0.062}_{-0.074}$ & $0.974^{+0.061}_{-0.066}$ & $0.924^{+0.060}_{-0.065}$  \\ 
\hline
\multirow{4}{*}{$I_c/I$ [\%]} & 1.0 & $7.5^{+1.0}_{-1.0}$ & $7.5^{+0.8}_{-0.9}$ & $6.1^{+0.7}_{-0.8}$  \\ 
  & 1.2 & $5.8^{+0.8}_{-0.8}$ & $5.8^{+0.6}_{-0.7}$ & $4.7^{+0.6}_{-0.6}$  \\ 
  & 1.4 & $4.6^{+0.7}_{-0.7}$ & $4.6^{+0.5}_{-0.6}$ & $3.7^{+0.5}_{-0.5}$  \\ 
  & 1.6 & $3.7^{+0.6}_{-0.5}$ & $3.7^{+0.4}_{-0.5}$ & $3.0^{+0.4}_{-0.4}$  \\ 
\hline
\end{tabular}
\end{table}

\subsection{Glitch activity and moment of inertia of the inner crust}

To account for the large glitch activity of the Vela pulsar (PSR J0835–4510), a substantial amount of angular momentum must be stored in the neutron star crust in the form of a superfluid neutron current~\cite{amp_review_2023}. 
This is possible only if the superfluid component in the crust carries a sufficiently large fraction of the star's moment of inertia~\cite{link1999,andersson_CNE_2012,Chamel2012,Montoli2021}. 
The estimated minimum required fractional moment of inertia strongly depends on the proportion of neutrons entrained by the normal component, represented by the ion lattice~\cite{Chamel2017JApA}; see also~\cite{amp_review_2023} for a review.
The strength of this entrainment phenomenon - which is an input for the hydrodynamic description of the inner crust~\cite{amp_review_2023,gavassino2021univ} - remains a subject of debate, as it depends on the nuclear interaction, subtle neutron band structure and pairing effects~\citep{Chamel2017JApA}. 
While a precise quantitative evaluation of the entrainment parameter is still under discussion - see~\cite{saluls2020arXiv,Minami_PRR_2022,Sekizawa2022,Chamel2024iej,Almirante2025arXiv} for recent developments - a minimal hypothesis is to assume that all and only the neutrons bound to the cluster are entrained (i.e., are locked to the normal component in the crust). 
This hypothesis closely resembles the negligible entrainment scenario suggested by the recent revision of the entrainment strength in the inner crust by~\citet{Almirante2025arXiv}.

The number of unbound neutrons in each Wigner-Seitz cell can be estimated from the neutron gas density $n_g$ as $N_{\text{unbound}} = n_g V_{WS}$, where $V_{WS}$ is the cell volume~\cite{Papakonstantinou2013}. 
This assumption, which we refer to as ``minimal entrainement'', defines an upper bound on the moment of inertia $I_n$ of the superfluid neutrons available to explain the large glitch activity of the Vela pulsar - that is, essentially, the early estimate of the moment of inertia fraction relevant for glitches proposed in the work of~\citet{link1999}. 

The dimensionless glitch activity parameter $\mathcal{G}$ quantifies, on average, the fraction of the star's spin-down that is reversed by glitches over long timescales~\cite{amp_review_2023}. 
For the Vela pulsar, different statistical methods yield a value of approximately $\mathcal{G}_{\text{Vela}} = 0.016 \pm 0.002$; see Table 1 in~\cite{Montoli2021}. 
It can be shown that the moment of inertia $I_n$ of the superfluid neutrons in the  region responsible for glitches (assumed to be the inner crust) must satisfy
\begin{equation}
    \frac{I_n}{I - I_n} > \mathcal{G} \, ,
    \label{eq:glitch_activity}
\end{equation}
where $I$ is the total moment of inertia of the star; see~\cite{amp_review_2023} for a discussion on how to include entrainment corrections in~$I_n$.

\begin{figure*}
    \subfloat[Spinodal]{
    \includegraphics[width=0.45\linewidth]{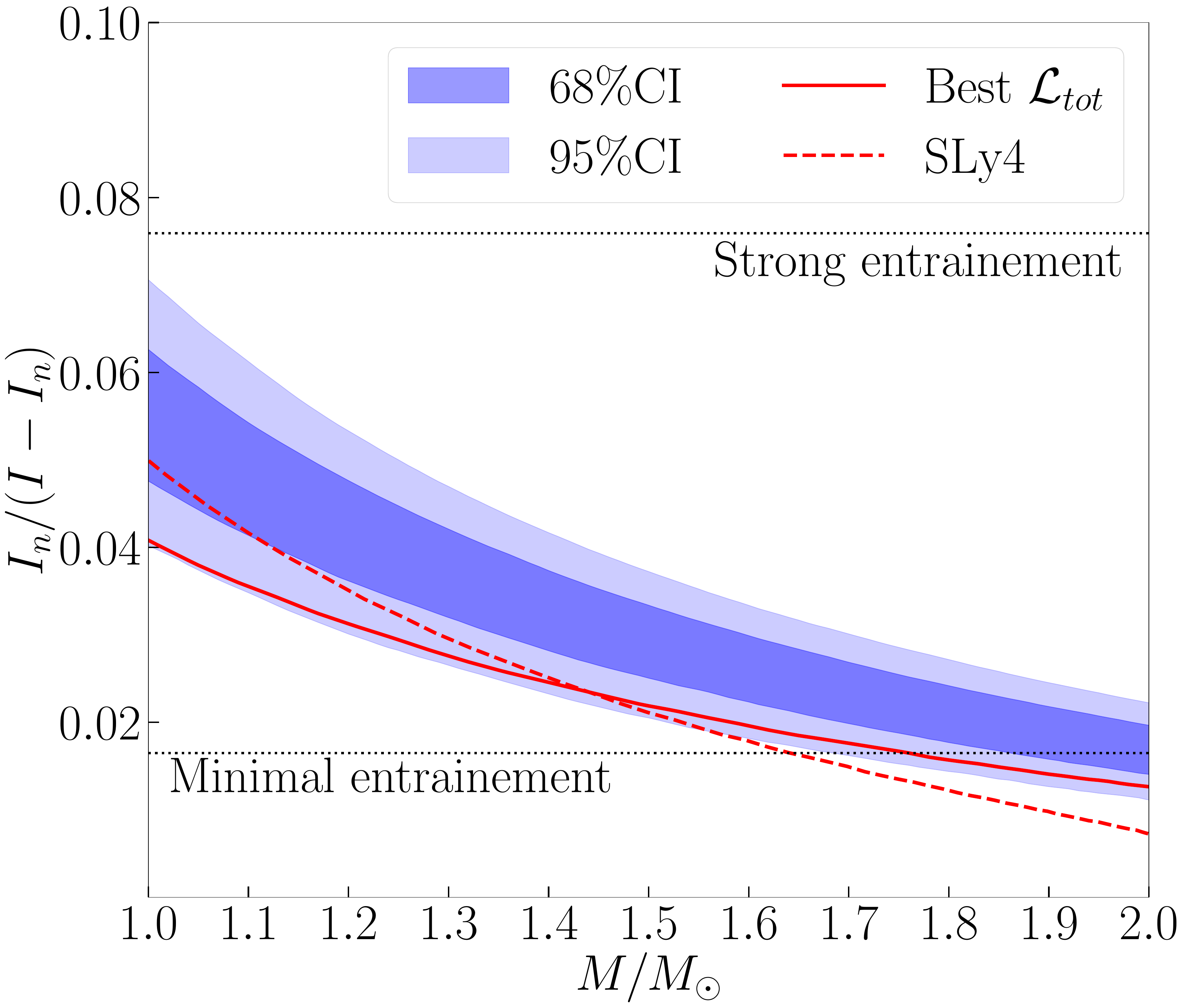}}
    \quad
    \subfloat[Linear]{
    \includegraphics[width=.45\linewidth]{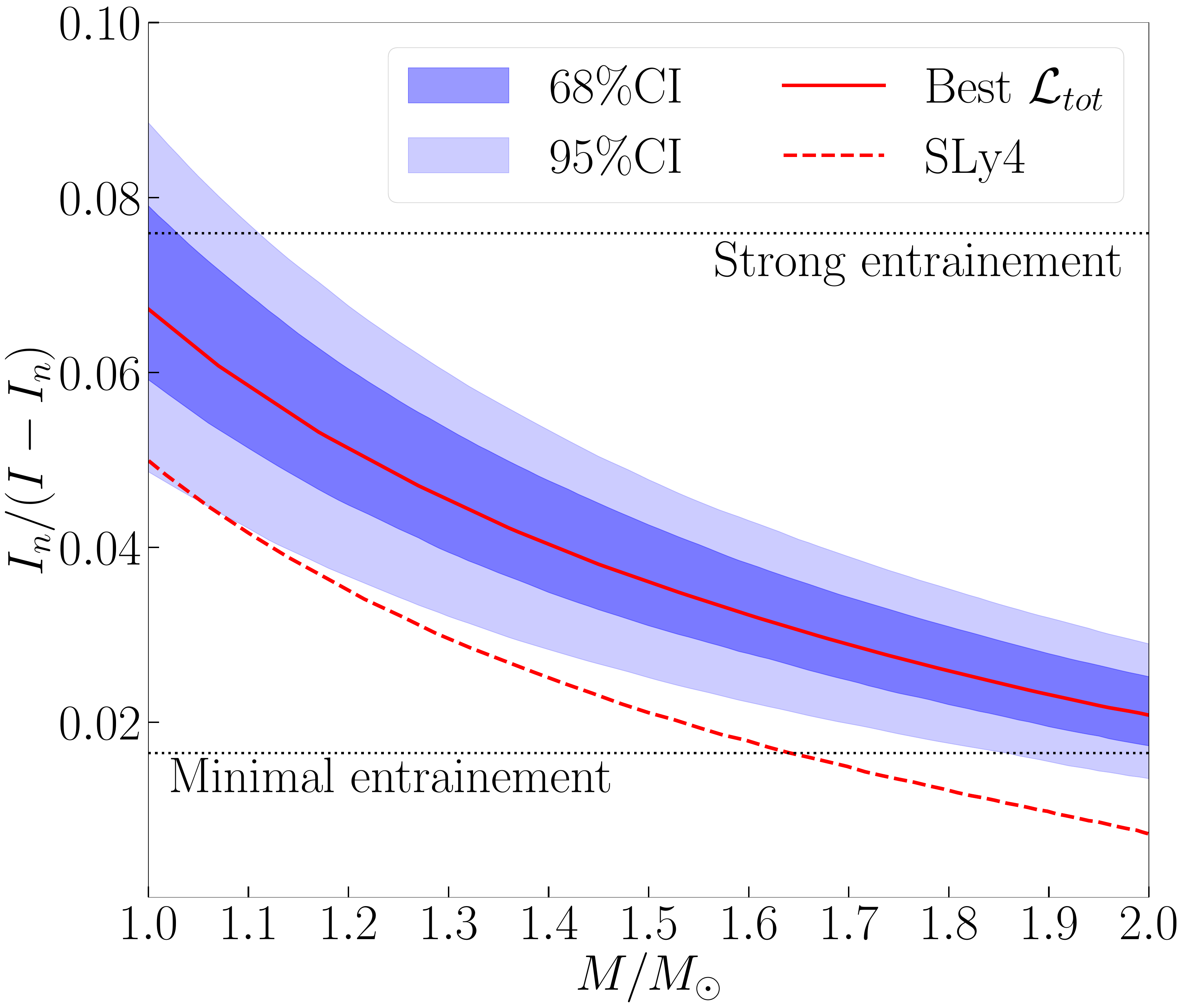}}
    \caption{Posterior of the ratio in~\eqref{eq:glitch_activity} as a function of the stellar mass, using the CC transition point defined via the dynamical spinodal (left) or the ETF crust calculation (right). The darker (lighter) blue bands represent the 68\% (95\%) confidence intervals. The solid (dashed) red line corresponds to the model with the highest $\lk_{tot}$ (the Sly4 model~\cite{Chabanat1997}). For comparison, the value $\mathcal{G}_{\text{Vela}}$ of Vela's glitch activity is shown with horizontal lines, in the cases of minimal~\cite{link1999} and strong~\cite{Chamel2012,andersson_CNE_2012,Montoli2021} entrainment.}
    \label{fig:glitch_activity}
\end{figure*}

In Figure~\ref{fig:glitch_activity} we show the posterior distribution of the quantity $I_n/(I-I_n)$ for different values of the star mass. 
If the zero entrainment hypothesis is retained, the glitch activity of the Vela pulsar is compatible with a crustal origin unless Vela's mass exceeds $\approx 2\,M_\odot$. 
Hence, Figure~\ref{fig:glitch_activity} can be seen as a Bayesian extension of the early result of~\cite{link1999}.

For comparison, in Figure~\ref{fig:glitch_activity} we also include the case corresponding to a scenario with strong entrainment, which can be approximated by multiplying the value of $\mathcal{G}_{\text{Vela}}$ by a prefactor\footnote{
    Multiplying $\mathcal{G}$ by a prefactor (equal to 1 in the case of zero entrainment) is a simple and practical approximation to estimate the potential impact of entrainment. However, this prefactor is not universal in principle: it can vary across different nuclear models and stellar masses. More accurate treatments, and the basis for this approximation, are discussed in~\cite{delsate2016,Montoli2021,amp_review_2023}.} 
of $\ approx 4.6$. This value reflects the effect of strong entrainment on the fraction of neutrons available to store angular momentum during a glitch, as previously considered in~\cite{Chamel2012,andersson_CNE_2012}.
In this strong entrainment case, the constraint~\eqref{eq:glitch_activity} can be satisfied only under the ETF definition of the CC transition point (panel b of Figure~\ref{fig:glitch_activity}), and only if the mass of the Vela pulsar is very low ($M \lesssim 1.1\,M_\odot$), in accordance with~\citep{Montoli2021}. 
Interestingly, under the less sophisticated crustal definition based on the spinodal instability criterion, the hypothesis of a crustal origin would be entirely ruled out in this case of strong entrainment.

\subsection{Mass-Radius relation and EoS}

\begin{figure*}
    \includegraphics[width=1\linewidth]{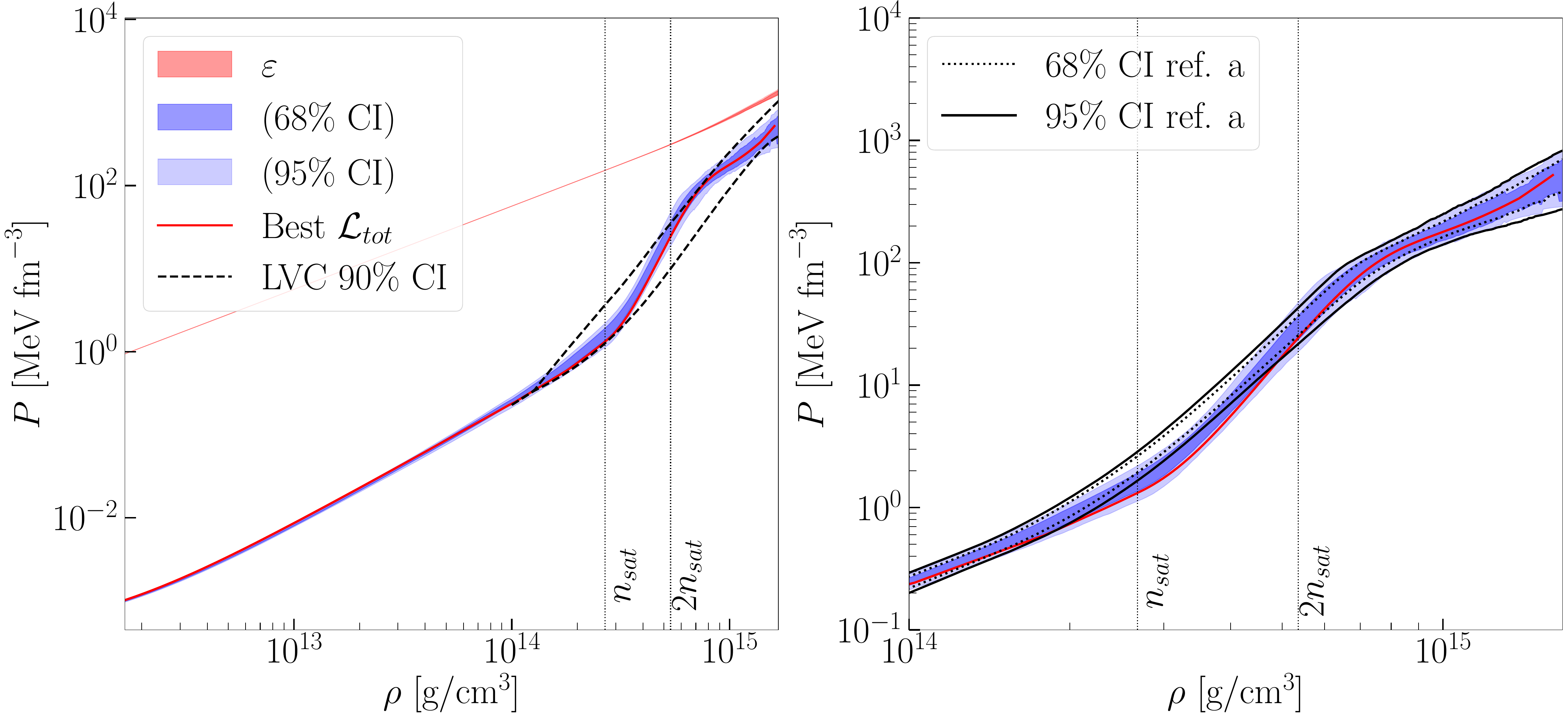}
    \caption{On the left, the posteriors of the pressure~$P$ (blue) and energy density~$\varepsilon$ (red) are shown as functions of the mass density~$\rho$. The darker shades indicate the 68\% confidence intervals, while the lighter shades correspond to the 95\%. 
    For comparison, the pressure of the best-fit model (highest~$\lk_{tot}$) is shown as a red line, and the dashed black region denotes the 90\% confidence interval from the GW170817 analysis~\cite{Abbott2018}. 
    On the right, a zoom of the pressure posterior in the crust. 
    This time, the black (dotted) line represents the (68\%) 95\% confidence interval of ref. a, \cite{DinhThi2021}.
    }
    \label{fig:eos_posterior}
\end{figure*}

We now turn to the posterior of the EoS and energy density of the star, shown in the left panel of Figure~\ref{fig:eos_posterior} as a function of the mass density~$\rho$. 
For comparison, the agnostic 90\% confidence interval from the GW170817 tidal polarizability measurement by the LVC~\cite{Abbott2018} is indicated by dashed lines.

\begin{figure*}
    \includegraphics[width=1\linewidth]{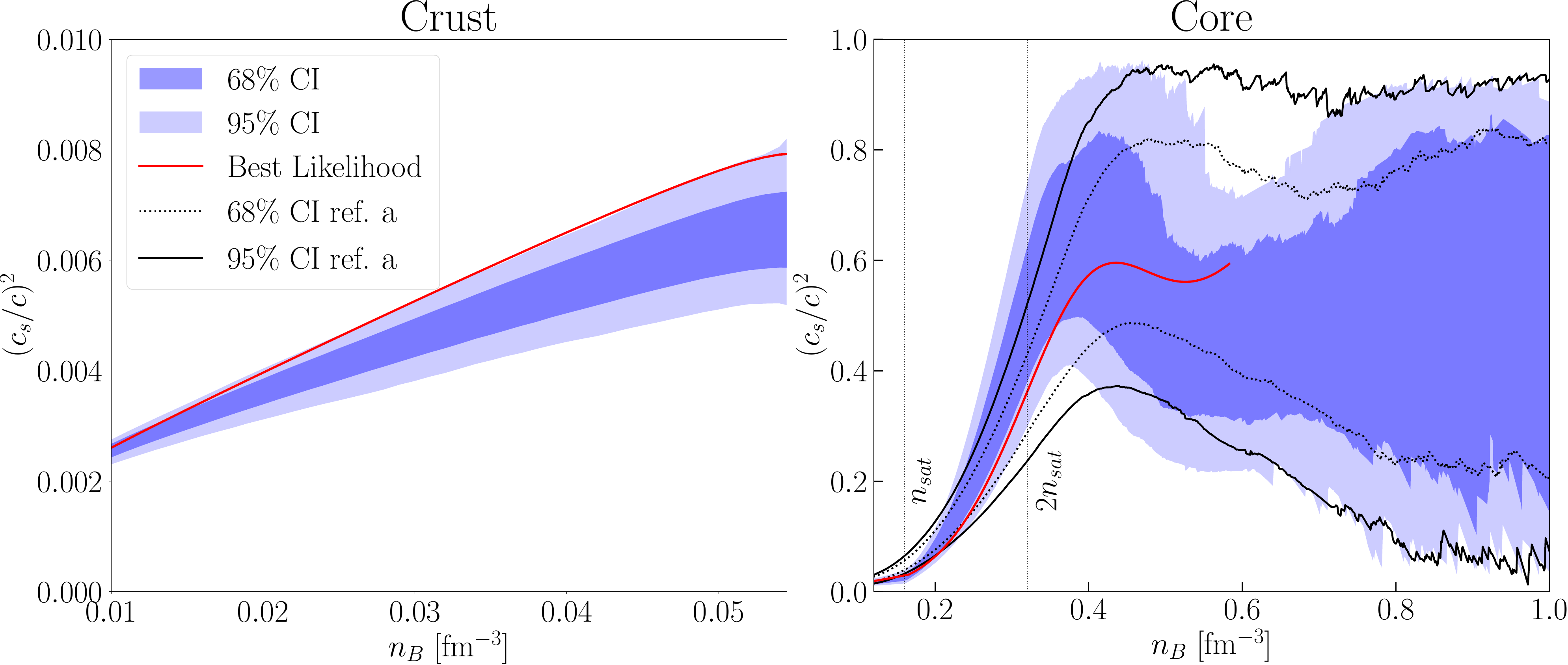}
    \caption{Posterior of the sound speed inside the star. The darker shades indicate the 68\% confidence intervals, while the lighter shades correspond to the 95\%. 
    The dotted and solid black lines delimit the same intervals for the result of ref. a, \cite{DinhThi2021}.
    For comparison, the pressure of the best-fit model (highest~$\lk_{tot}$) is shown as a red line.
    The left subplot focuses on the inner crust, while the right on the core.}
    \label{fig:sound_speed}
\end{figure*}

The posterior of the energy density~$\varepsilon$ is shown in red, with the darker shade marking the 68\% confidence region and the lighter shade the 95\%. 
Overall, it exhibits very weak dependence on the nuclear interaction, as evident from the narrow width of the distribution. 
At very high densities ($\ approx 5\,\nsat$), the spread increases slightly, though not significantly. 
The pressure~$P$ posterior is plotted in blue, while the model yielding the highest~$\lk_{tot}$ is indicated by the red line. 
The overall distribution is in good agreement with the agnostic LVC posterior but considerably thinner, thus highlighting the important information brought by nuclear physics theory and experiments.
In the present work, only results from ab-initio nuclear theory and nuclear structure experiments are considered, which limits the laboratory information to relatively low densities.
A more precise determination of the high-density behavior $n_B\gtrsim 2n_{sat}$ would require constraints from heavy-ion collision data~\cite{sorensen}.
The specific information brought by nuclear structure data can be appreciated from the right panel of Figure~\ref{fig:eos_posterior}, which shows a zoom of the EoS and a comparison with the posterior probability distribution found in~\cite{DinhThi2021}. 
In that paper, the authors applied the same metamodel treatment of the EoS and the same chiral and astrophysical constraints as in the present work, but they employed a CLDM treatment for the crust and an agnostic prior distribution for the bulk nuclear matter parameters and effective masses. 
The comparison between the two posterior distributions thus quantifies how the nuclear data information embedded in the analysis of \cite{Klausner2025} propagates into the knowledge of the neutron star EoS. 
We can see that in the present work, star matter is considerably soft near saturation, reflecting the overall low~$\lsym$ values predicted by the analysis (see Figure~\ref{fig:marg_posteriors}). Beyond~$\nsat$, the EoS stiffens rapidly to support the maximum observed mass, and then softens again to coincide with the results of \cite{DinhThi2021} at higher densities, where only general physical constraints such as stability and causality govern the metamodel predictions.
The right panel of Figure \ref{fig:eos_posterior} focuses on the pressure posterior in the star core.
This time, we compare our findings with the results of \cite{DinhThi2021}, whose 68\% and 95\% confidence intervals are delimited with dashed and continuous black lines.
The stiffening of the equation of state is evident between one and two saturation densities.
In Figure \ref{fig:sound_speed} we show the squared sound speed\footnote{
    To keep the comparison with \cite{DinhThi2021}, we consider the barotropic sound speed rather than the adiabatic (or frozen) one~\cite{Montefusco2025}: since the barotropic sound speed coincides with the one calculated by considering the TOV energy-pressure profile, this is the quantity that measures the stiffness of the chemically equilibrated EoS; see, e.g., the discussion in~\cite{giliberti2020MNRAS}. 
} $(c_s/c)^2$, again comparing it to the results of \cite{DinhThi2021}.  
We observe that $(c_s/c)^2$ increases rapidly between one and two $\nsat$ $c_s$: this rapid growth corresponds to the stiffening of the equation of state in this density region. 

\begin{figure}
    \includegraphics[width=1\linewidth]{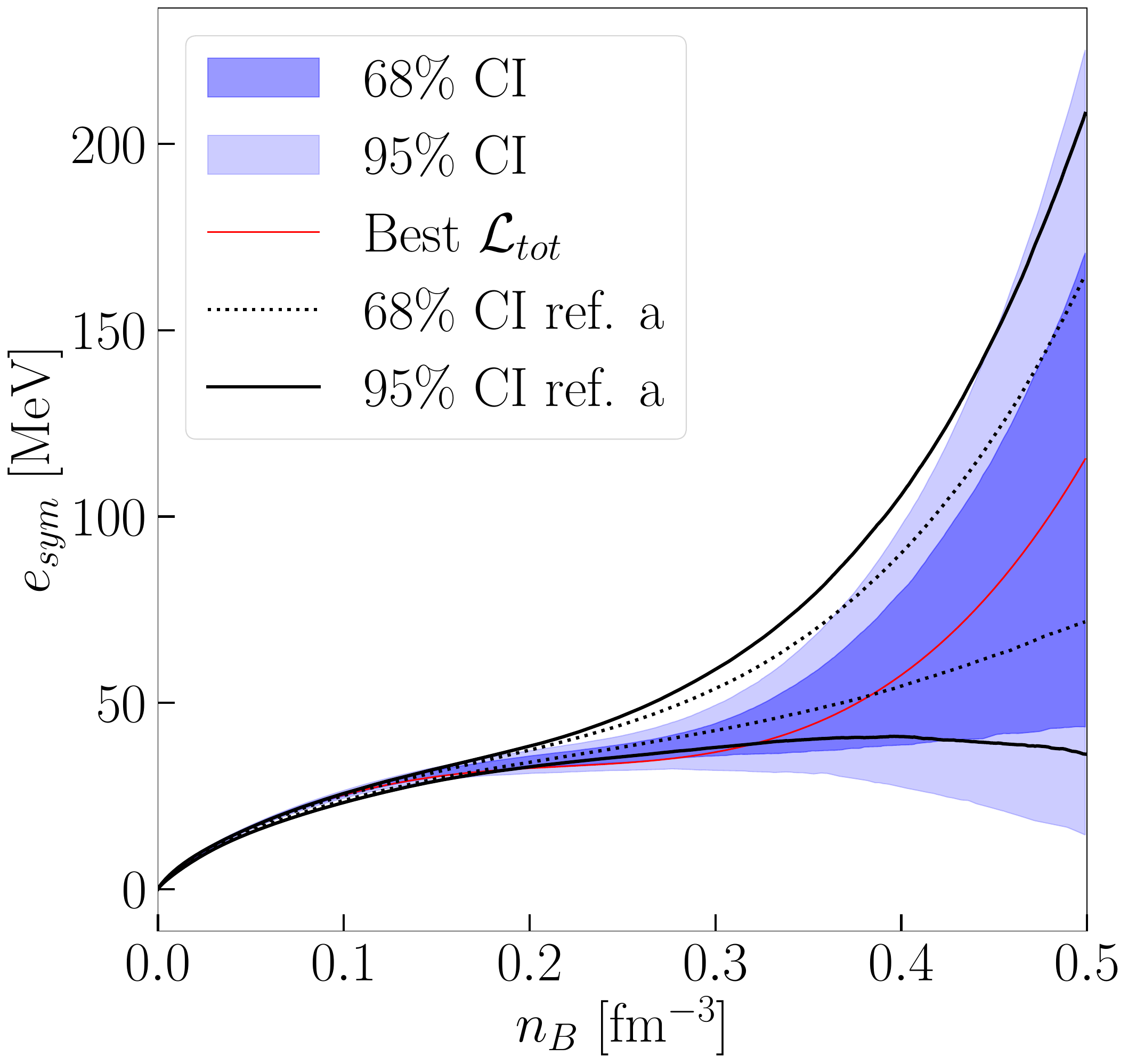}
    \caption{Symmetry energy posterior.
    The darker shades indicate the 68\% confidence intervals, while the lighter shades correspond to the 95\%. 
    The symmetry energy of the best-fit model is in red.
    The black (dotted) line represents the (68\%) 95\% confidence interval of ref. a,~\cite{DinhThi2021}.}
    \label{fig:esym_posterior}
\end{figure}

Figure~\ref{fig:esym_posterior} displays the posterior of the symmetry energy. 
As before, the distribution is narrow up to~$\nsat$, and broadens significantly at higher densities.
The distribution is again compatible with the results of \cite{DinhThi2021}, but we can again appreciate the softening of the symmetry energy as imposed by the nuclear structure-informed prior used in this work. 
At very high-density, where no nuclear physics information is available, our predictions are slightly more spread than in \cite{DinhThi2021}. 
This might be because we use independent high-order parameters ($Q_{sat,sym}$ and $Z_{sat,sym}$) above saturation, thus completely decoupling the low-density information from the one at high-density, where additional degrees of freedom might pop up and the Taylor expansion becomes meaningless. 

\begin{figure}
    \includegraphics[width=1\linewidth]{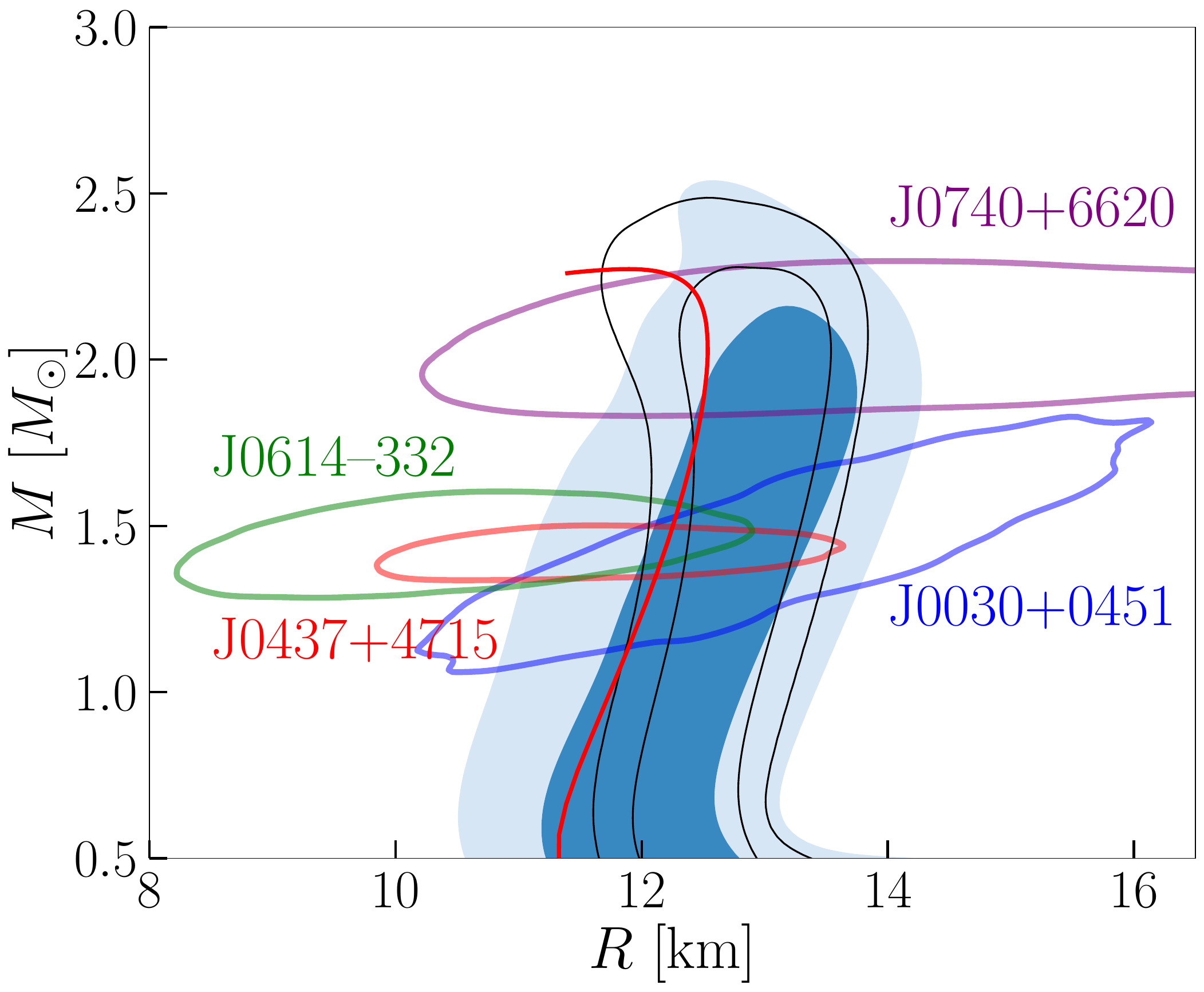}
    \caption{Posterior of the MR relation. 
             The darker blue encloses the 68\% probability region, while the lighter blue encloses the 95\%.
             The five best models are highlighted in red. 
             The four colored contours enclose the 95\% probability regions of the four NICER measurements used in our analysis.
             The regions in the dark lines represent the posterior probability distribution found in~\cite{DinhThi2021} (68\% the smallest, 95\% the bigger).
    }
    \label{fig:MR}
\end{figure}

We conclude our results section with the mass-radius plot shown in Figure~\ref{fig:MR}.  
For reference, the five models corresponding to the highest values of~$\lk_{tot}$ are displayed. 
The two black contours indicate the 68\% confidence regions of the NICER inferences included in our analysis. 
Instead, the orange-shaded regions represent the posterior probability distribution found in~\cite{DinhThi2021}.
The different colors delimit the regions containing, from darker to lighter, the 68\% and 95\% of the distribution.
The results are compatible, especially at high NS masses ($M\ approx 2M_\odot$), where our results are well within the 68\% region.
On the other hand, at low NS masses ($M< 1.5M_\odot$), our analysis predicts thinner distributions and a non-negligible shift towards smaller radii, as a clear consequence of the softer symmetry energy. 

\section{Conclusions}
\label{sec:concl}

We performed a Bayesian analysis of the neutron star EoS, combining constraints from the maximum observed neutron star mass, the tidal deformability measurement from GW170817, and simultaneous mass-radius inferences from NICER. 
To further refine the treatment of the inner crust and the outer core - where the neutron gas plays a crucial role - we also incorporated input from ab initio calculations of neutron matter at low densities, specifically the energy band predicted by different ab-initio calculations of pure neutron matter from \cite{Huth2021}.

Regarding the Bayesian framework, the novelty of this work lies in the integration of the recent nuclear physics posteriors of \citet{Klausner2025}, which are included directly at the level of our prior. 
In this way, we ensured that our final results reflect the combined influence of terrestrial nuclear measurements in a more refined way than previous similar studies - e.g., \cite{DinhThi2021,Pradhan_2023,Char_2023,Malik_2024,Montefusco2025} - based on the independent sampling of nuclear matter parameters: by using the posterior of \cite{Klausner2025} as our prior, we can (i) effectively incorporate the effect of the nuclear surface properties in the determination of the nuclear matter parameters of interest, and (ii) account for all their mutual correlations.

A further improvement over previous Bayesian studies relies in the nuclear structure informed unified Bayesian modelling of the crust, incorporating the correlated bulk and surface parameters posterior of \citet{Klausner2025} for an improved estimation of the crust-core transition point.
This latter is obtained both from the onset of the dynamical spinodal instability and from the more microscopic ETF variational method, though with some simplifications imposed by the complexity of the numerical effort. 
We find that the ETF method leads to an increased average value for the estimated CC transition pressure of 12\% with respect to the spinodal criterion, that in turn considerably affects the predictions for the crustal radius and moment of inertia.
 
The most important effect of the nuclear structure informed prior is an increased softness of the NS EoS around saturation, leading to a sizeable concavity in the region $n_{sat}\lesssim n_B\lesssim 2 n_{sat}$. 
This finding is strongly correlated to the soft symmetry energy behavior reported in~\cite{Klausner2025} as due to the combined inference of static and dynamic nuclear structure properties.
While this very soft symmetry energy behavior is still compatible at $2\sigma$ with the general systematics from terrestrial experiments of ref.~\cite{BaoAn_2021}, further investigation is needed to assess the possible influence of the choice of nuclear data used in~\cite{Klausner2025}. 
Sensitivity studies on prior distributions would be valuable to assess the robustness of these conclusions.

From the nuclear physics viewpoint, our findings highlight the non-negligible impact of astrophysical constraints on saturation properties of nuclear matter, as a complementary source of information with respect to nuclear physics experiments.
In particular, the symmetry energy at saturation ($E_{sym}$) and its slope parameter ($L_{sym}$) tend to shift toward higher values with respect to the earlier analysis of~\cite{Klausner2025}, when observational constraints are applied. 

In conclusion, we believe that our results are relevant for crust modelling and for assessing the uncertainties in its composition and structure, where microphysical input consistent with both laboratory experiments and ab initio calculations is essential. 
Future work should aim to reduce systematic uncertainties, for instance by improving the treatment of the crust-core transition and incorporating additional experimental nuclear data, notably on open-shell and exotic nuclei, through an even more refined prior.

\section*{ACKNOWLEDGMENTS}
We are thankful to H.Dinh-Thi for providing the data files of the results published in ref.~\cite{DinhThi2021} for the comparisons of section \ref{sec:res}, to T.Diverres and A.F.Fantina for sharing with us the ETF results using the code of ref.~\cite{Pearson2018}, and to G.Montefusco for sharing parts of the code developed for ref.~\cite{Montefusco2025}.
F.G. and M.A. acknowledge the support by the IN2P3 Master Project NewMAC and MAC, the ANR project `Gravitational waves from hot neutron stars and properties of ultra-dense matter' (GW-HNS, ANR-22-CE31-0001-01), the CNRS International Research Project (IRP) `Origine des \'el\'ements lourds dans l'univers: Astres Compacts et Nucl\'eosynth\`ese (ACNu)'.

\bibliography{bibliography}

\newpage
\appendix

\section{ETF crust modelling}
\label{app:crust}
\begin{figure*}
    \includegraphics[width=0.85\linewidth]{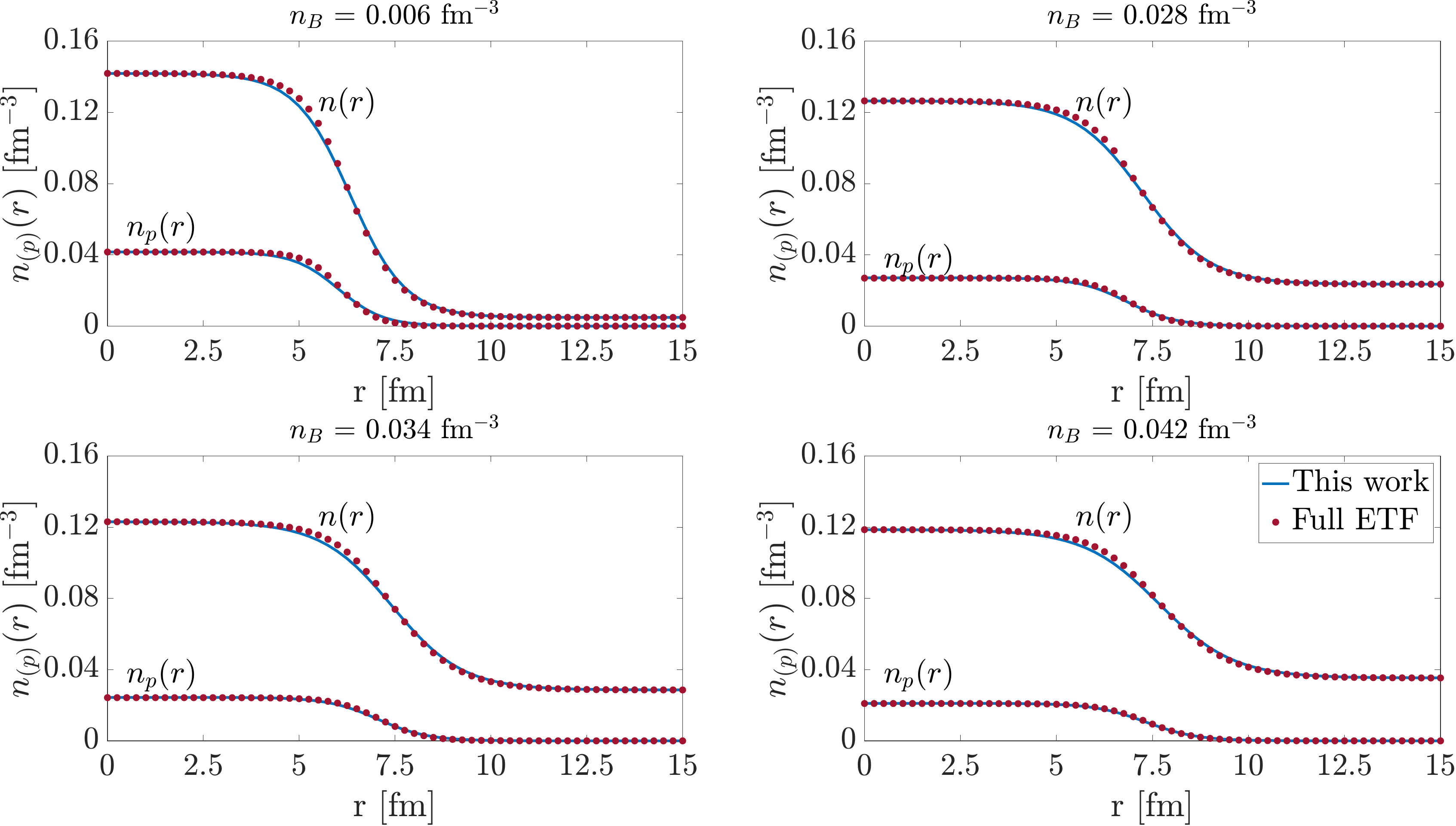}
    \caption{Comparison of the optimal Sly4 density profiles in the full ETF variation of ref.\cite{Pearson2018} and the ones using approximate values for $n_{\rm 0p},a,a_{\rm p}$ from Eqs.(\ref{eq:delta_eff}),(\ref{eq:a_from_delta}).}
    \label{fig:density_parametrization_comparison}
\end{figure*}

\begin{figure*}
    \includegraphics[width=0.85\linewidth]{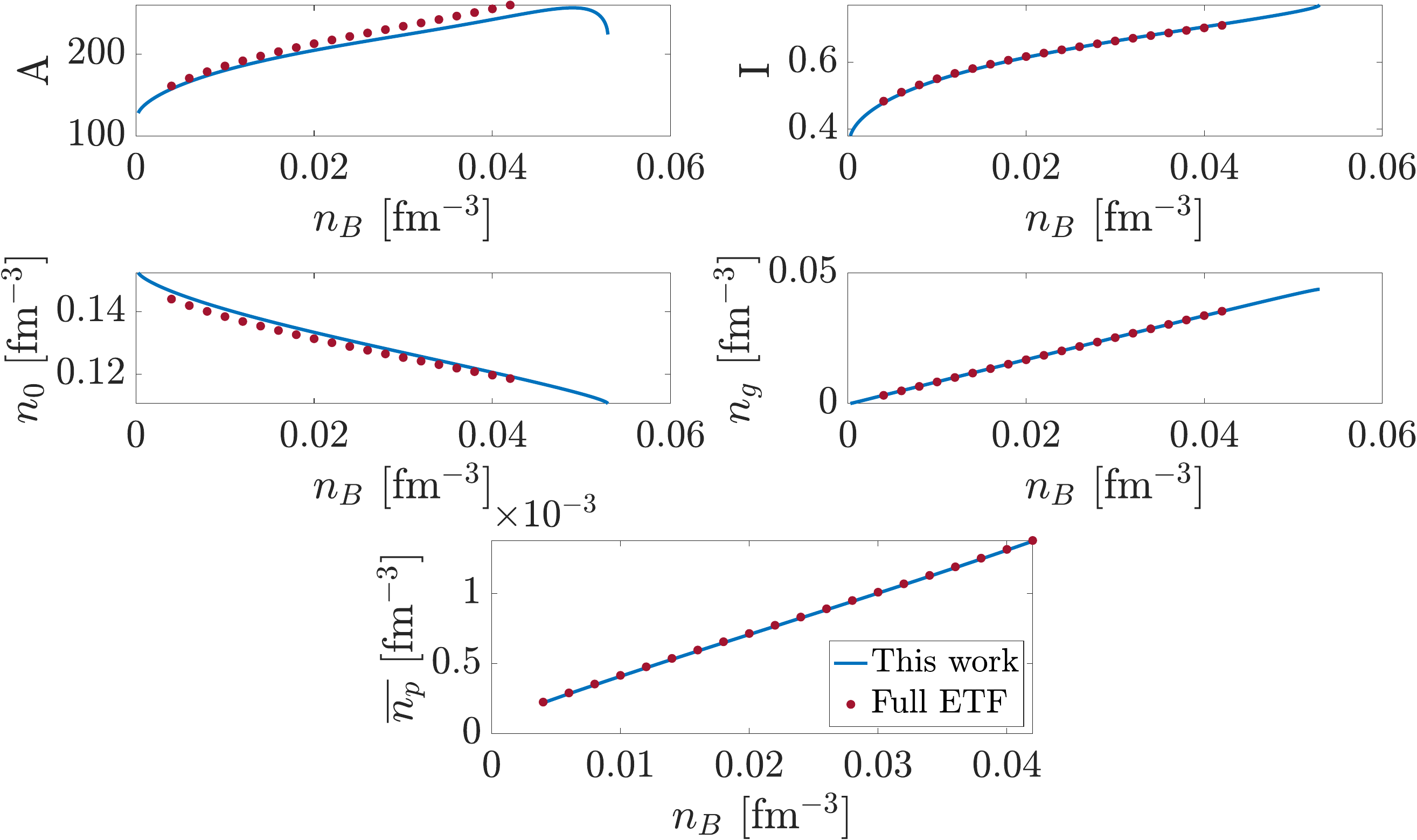}
    \caption{ Comparison of the cell variational parameters obtained in this work, with the ones from ref.\cite{Pearson2018}, in the case of the Sly4 interaction.}
    \label{fig:crust_compo_comparison}
\end{figure*}

 The energy density of the Wigner-Seitz cell, $\varepsilon_{WS}$, reads
\begin{equation}
    \varepsilon_{WS} = \frac{1}{V_{WS}}\int_{V_{WS}} \mathcal{E}_{ETF}^{Sky}(n(\textbf{r}), n_{\rm p}(\textbf{r})) d^3\textbf{r} +\varepsilon_e.
    \label{eq:cell_en_dens}
\end{equation}
In this expression, $\varepsilon_e$ is the electron energy density
\begin{equation}
    \varepsilon_e = \frac{c}{\pi^2} \int_{0}^{k_{F_e}} \!\! d k \, k^2 
    \sqrt{\hbar^2 k^2 + m_e^2 c^2},
    \label{eq:en_dens_el}
\end{equation}
where the electron Fermi momentum is related to the electron density $n_{\rm e}$ by $k_{F_e}=(3\pi^2n_{\rm e})^{1/3}$, and charge neutrality imposes $n_e=\bar{n}_p$, $\bar{n}_p$ being the average proton density in the cell. 
The Skyrme energy density in the extended Thomas-Fermi approximation can be decomposed as
\begin{equation}
  \mathcal{E}_{ETF}^{Sky} = \mathcal{E}_{\rm kin} +\mathcal{E}_{\rm int}+\mathcal{E}_{\rm Coul},
\label{eq:edf}
\end{equation}
where the Coulomb and interaction terms are unchanged with respect to the full Skyrme form \footnote{The isoscalar $t=0$ and isovector $t=1$ densities are defined as a function of the local baryon and proton densities as $n_0=n$, $n_1=n-2n_{\rm p}$} \cite{Pearson_2020}:
\begin{eqnarray}
  \mathcal{E}_{\rm int}&=& \sum_{t=0,1} C_t^\rho n_t^2+C_t^{\Delta \rho}n_t\Delta n_t\nonumber\\&+&C_t^\tau n_t\tau_t+\frac{1}{2}C_t^J{\bm J}_t^2+C_t^{\bm\nabla J}n_t{\bm\nabla}\cdot{\bm J}_t,\\
  \mathcal{E}_{\rm Coul}&=&\frac{8\pi e^2}{V_{\rm WS}}\int_0^{R_{\rm WS}} \frac{dr}{r^2}\left( \int_0^r\left (n_{\rm p}(r^\prime)-n_{\rm e}\right )r^{\prime 2} dr^\prime\right)^2\nonumber
   ,
  \label{eq:skyrme_functional}
\end{eqnarray}
and the ETF approximation allows expressing the non-local operators as a function of the local densities $n(r), n_{\rm p}(r)$. 
We have for the kinetic energy density expanded to second order in $\hbar$ :
\begin{equation}
    \tau=\sum_{\rm{q=n,p}} \frac{3}{5}\left(3 \pi^2\right)^{2 / 3}  n_{\rm{q}}{5/3}+ \tau_{q, 2}^L+\tau_{q, 2}^{N L}
\end{equation}
where,
\begin{equation}
\begin{aligned}
\tau_{q, 2}^L= & \frac{1}{36} \frac{\left(\nabla n_q\right)^2}{n_q}+\frac{1}{3} \Delta n_q, \\
\tau_{q, 2}^{N L}= & \frac{1}{6} \frac{\nabla n_q \cdot \nabla f_q}{f_q}+\frac{1}{6} n_q \frac{\Delta f_q}{f_q}-\frac{1}{12} n_q\left(\frac{\nabla f_q}{f_q}\right)^2 \\
& +\frac{1}{2}\left(\frac{2 m}{\hbar^2}\right)^2 n_q\left(\frac{W_q}{f_q}\right)^2 ,
\end{aligned}
\end{equation}
 and the effective mass term $f_q=m / m_q^*$ reads
\begin{equation}
f_q=1+\frac{2 m}{\hbar^2}\left[\left(C_0^\tau+C_1^\tau\right) n_q+\left(C_0^\tau-C_1^\tau\right) n_{\bar{q}}\right] .
\end{equation}

The spin-orbit current obtained at the same $\hbar^2$ order in the semi-classical expansion is given by~\cite{Brack1985} 
\begin{equation}
J_q=-\frac{2 m}{\hbar^2 f_q} n_q W_q,
\end{equation}
where the spin-orbit potential $W_q$ reads
\begin{equation}
\begin{aligned}
W_q= & -\left(C_0^{\nabla J}+C_1^{\nabla J}\right) \nabla n_q-\left(C_0^{\nabla J}-C_1^{\nabla J}\right) \nabla n_{\bar{q}} \\
& +\left(C_0^J+C_1^J\right) J_q+\left(C_0^J-C_1^J\right) J_{\bar{q}} .
\end{aligned}
\end{equation}

The relation between the spin currents $J_n$ and $J_p$ and the gradient of the densities is thus given by the solution of the $2 \times 2$ system of linear equations
\begin{equation}
\begin{aligned}
& \left(\frac{\hbar^2}{2 m} f_q+\left(C_0^J+C_1^J\right) n_q\right) J_q+\left(C_0^J-C_1^J\right) n_q J_{\bar{q}} \\
& \quad=\left(C_0^{\nabla J}+C_1^{\nabla J}\right) n_q \nabla n_q+\left(C_0^{\nabla J}-C_1^{\nabla J}\right) n_q \nabla n_{\bar{q}} .
\end{aligned}
\end{equation}
We use parametrized Woods-Saxon profiles 
for the local densities:
\begin{equation}
\begin{aligned}
    n(r)   =& \frac{n_0}{1 + \exp \left(\frac{r - R}{a}\right)} + 
     \frac{n_g}{1 + \exp \left(-\frac{r - R}{a}\right)} \\
    n_p(r)  =& \frac{n_{0,\,p}}{1 + \exp \left(\frac{r - R_p}{a_p}\right)},
    \label{eq:density_profiles}
\end{aligned}
\end{equation}
such that the ETF problem amounts to minimizing the energy density Eq.(\ref{eq:cell_en_dens}) with respect to 8 independent variables: $n_g,n_0,n_{0p},R,R_p,a,a_p,V_{WS}$.
Such a minimization problem is very numerically demanding for the Bayesian study of this work, where the crust has to be computed typically for some $10^4$ different models to get convergent results. 
The number of independent variables can be reduced if we assume that the bulk isospin asymmetry $\delta=(n_0-2n_{\rm 0p})/n_0$ can be obtained from the pressure equilibrium condition between a bulk matter at density $n_0$ and a homogeneous neutron gas at density $n_g$. 
Such a condition reads:
\begin{equation}
    n^2\left.\frac{\partial (\varepsilon_B(n,\delta)/n)}{\partial n}\right|_{n_0} = P_g(n_g, \delta_g = 1)
    \label{eq:delta_relation}
\end{equation}
where $\varepsilon_B$ is the energy density of homogeneous nuclear matter, and $P_g$ is the pressure of the pure neutron gas. Since the equilibrium bulk density $n_0$ is expected to be close to the saturation density $n_{sat}$ of symmetric nuclear matter~\cite{Pearson_2020}, in Eq.(\ref{eq:delta_relation}) the function $\varepsilon_B$ can be expanded around this point giving an analytical expression of $\delta$ as a function of the nuclear matter parameters $K_{sat},L_{sym},K_{sym}$: 
\begin{equation}
    \delta =   \sqrt{\frac{(n_{sat}-n_0)K_{sat} + \left(3n_{sat}/n_0\right)^2 P_g}{3L_{sym}n_{sat} - K_{sym}(n_{sat}-n_0)}}.
    \label{eq:delta_eff}
\end{equation}
An extra reduction of the computational effort can be achieved by using the results of \cite{Papakonstantinou2013}, where the authors performed extensive Woods-Saxon fits of Hartree-Fock (HF) profiles in the density and asymmetry conditions typical of the inner crust. 
They showed that the diffuseness parameters $a,a_{\rm p}$ follow an approximate quadratic law with the bulk asymmetry, and proposed a quasi-universal parametrization with an optimal fit:
\begin{equation}
\begin{aligned}
    a   &= 0.54 + 1.04\times\delta^2 \\
    a_{\rm p} &= 0.53 + 0.33\times\delta^2.
    \label{eq:a_from_delta}
\end{aligned}
\end{equation}
Using Eqs.(\ref{eq:delta_eff}),(\ref{eq:a_from_delta}), the ETF variational problem is reduced to a dimensionality $N=5$, which is the same numerical cost as the CLDM minimization. 
Moreover, a one-to-one mapping can be established between the ETF residual variational variables $(n_0,n_g,R,R_{\rm p},V_{WS})$ and the CLDM variational variables $(n_0,n_g,A,I,\bar{n}_{\rm p})$ used in~\cite{DinhThi2021b}. 
This allowed us to use the same numerical code to compute the crust both in the CLDM and ETF formalisms, thus minimizing the technical problems that arise when different numerical structures are interfaced.

Specifically, the average proton density relates to the cell volume and the average cluster baryon number $A$ and asymmetry $I$ via $\overline{n}_{\rm p} = Z / V_{WS} = A(1 - I)/2 / V_{WS}$.
Concerning the radius parameters, they are mapped into effective CLDM-like variational variables through the following relations \cite{Warda2010}:
\begin{equation}
    R_{({\rm p})} = R_{HS(,\,{\rm p})} \left(1 - \frac{\pi^2}{3} \left(\frac{a_{({\rm p})}}{R_{HS(,\,{\rm p})} }\right)^2\right),
    \label{eq:wood_saxon_radius_parameters}
\end{equation}
where the hard-sphere radii are defined as  \footnote{ These radii define uniform spheres containing the same number of nucleons (or protons) as the diffused distribution and with the corresponding central density.} 
\begin{equation}
    \begin{aligned}
        R_{HS} &= \left(\frac{3}{4\pi}\frac{A}{n_0}\right)^{1/3}, \\
        R_{HS,\,p} &= \left(\frac{3}{4\pi}\frac{(1 - I)A}{2n_{0,\,p}}\right)^{1/3}.
        \label{eq:r_hard_sphere}
    \end{aligned}
\end{equation}

The minimization of the energy density Eq.(\ref{eq:cell_en_dens}) with respect to the reduced variable set $(n_0,n_g,R,R_{\rm p},V_{WS})$ (or the equivalent CLDM-like parameters $(n_0,n_g,A,I,\bar{n}_{\rm p})$ defined by Eq.(\ref{eq:wood_saxon_radius_parameters})) leads to the optimal ETF profile only if the ansatz Eqs.(\ref{eq:delta_eff}),(\ref{eq:a_from_delta}) are good approximations of the optimal values of $n_{\rm 0p},a,a_{\rm p}$.

We tested the validity of these approximations in two ways, employing as a benchmark the vastly tested SLy4 parametrization. 
First, we used the optimal profiles at various baryon densities from a well-tested ETF code for the crust~\cite{Pearson2018}, which minimizes the energy of the cell using exactly our same ansatz for the parametrized density profiles. 
By inverting Eqs.(\ref{eq:delta_eff}),(\ref{eq:a_from_delta}) we extracted our model estimations of $n_{\rm 0p},a,a_{\rm p}$ and 
obtained the approximate solutions for the profiles. 
The results for the model Sly4 at some representative densities in the inner crust are shown in Figure~\ref{fig:density_parametrization_comparison}. 
The very satisfactory overall agreement shows that Eqs.(\ref{eq:delta_eff}),(\ref{eq:a_from_delta}) give good estimations of $n_{\rm 0p},a,a_{\rm p}$  in the case of the optimal profile.

A stronger validation test consists of the direct comparison of the results of our minimization with those of ref.\cite{Pearson2018}.   
Owing to the different numerical methods involved, larger discrepancies are expected; nonetheless, a general satisfactory consistency is observed. 
Figure~\ref{fig:crust_compo_comparison} shows the resulting optimal cell compositions, including $A$, $Z$, and related quantities, again for the benchmark comparison of the Sly4 model.

\section{Sampled Skyrme} \label{app:sampled_skyrme}

We present in Table~\ref{tab:skmsmpl} the parameters of the three sampled Skyrme SkmSmpl1 and SkmSmpl2 used throughout the text.

\begin{table}
\centering
\caption{Skyrme parameters of the two sampled Skyrme SkmSmpl1 and SkmSmpl2 used throughout the text.}
\label{tab:skmsmpl}
\begin{tabular}{llcc}
\hline
      &     &   SkmSmpl1    &   SkmSmpl2    \\ 
\hline
$t_0$ & [MeV fm$^3$]& -1992.157 & -1983.031 \\ 
$t_1$ & [MeV fm$^5$]& 440.755   & 334.366   \\ 
$t_2$ & [MeV fm$^5$]& 96.649    & -124.556  \\ 
$t_3$ & [MeV fm$^{3+3\alpha}$]& 11910.389  & 13066.227 \\
$x_0$ & [-] & 0.439     &   0.385 \\ 
$x_1$ & [-] & -0.708 &0.197 \\ 
$x_2$ & [-] &  -2.999 & 0.334 \\ 
$x_3$ & [-] & 0.616 & 0.315 \\ 
$\alpha$ & [-] & 0.256 & 0.276 \\ 
$w_0$ & [MeV fm$^5$] & 144.399 &  116.148 \\ 
\hline
\end{tabular}
\end{table}




\end{document}